\documentclass[a4paper,11pt]{article}
\pdfoutput=1 

\usepackage{jheppub}
\usepackage{graphicx}
\usepackage{epsfig}
\usepackage{amsmath, amsthm, amssymb}
\usepackage{amsfonts}
\usepackage{subfigure}
\usepackage{color}
\usepackage{xspace}
\usepackage[dvipsnames]{xcolor}
\usepackage{paralist}
\usepackage{multirow}

\newcommand{\beq}{\begin{equation}}
\newcommand{\eeq}{\end{equation}}
\newcommand{\bea}{\begin{eqnarray}}
\newcommand{\eea}{\end{eqnarray}}
\newcommand{\bfig}{\begin{figure}}
\newcommand{\efig}{\end{figure}}
\newcommand{\bc}{\begin{center}}
\newcommand{\ec}{\end{center}}

\def\sq2{\sqrt{2}}
\newcommand{\gr}[1]{\ensuremath{\mathcal{#1}}\xspace}

\newcommand{\PZprime}{\ensuremath{Z'}\xspace}
\newcommand{\PZ}{\ensuremath{Z}\xspace}
\newcommand{\Pphoton}{\ensuremath{\gamma}\xspace}

\newcommand{\p}[1]{\ensuremath{#1^{\prime}}\xspace}

\newcommand{\Pq}{{q}}
\newcommand{\Pqt}{{t}}

\newcommand{\MSb}{\ensuremath{\overline{\mathrm{MS}}}\xspace}

\preprint{LPSC-15-318, MS-TP-15-25, SMU-HEP-15-13}
\title{Electroweak top-quark pair production at the LHC with $Z'$ bosons to NLO QCD in POWHEG}

\author[a]{Roberto Bonciani,}
\author[b]{Tom\'{a}\v{s} Je\v{z}o,}
\author[c]{Michael Klasen,}
\author[d]{Florian Lyonnet,}
\author[e]{Ingo Schienbein}

\affiliation[a]{Dipartimento di Fisica, Universit\`a di Roma ``La Sapienza'' and
INFN, Sezione di Roma, I-00185, Roma, Italy}

\affiliation[b]{Universit\`a di Milano-Bicocca and INFN, Sezione di Milano-Bicocca,
 Piazza della Scienza 3, I-20126 Milano, Italy}

 \affiliation[c]{Institut f\"ur Theoretische Physik, Westf\"alische
 Wilhelms-Universit\"at M\"unster, Wilhelm-Klemm-Stra\ss{}e 9, D-48149
 M\"unster, Germany}

\affiliation[d]{Southern Methodist University, Dallas, TX 75275, USA}

\affiliation[e]{Laboratoire de Physique Subatomique et de Cosmologie,
 Universit\'e Joseph Fourier/CNRS-IN2P3/ INPG,
 53 Avenue des Martyrs, F-38026 Grenoble, France}

\emailAdd{roberto.bonciani@roma1.infn.it}
\emailAdd{tomas.jezo@mib.infn.it}
\emailAdd{michael.klasen@uni-muenster.de}
\emailAdd{flyonnet@smu.edu}
\emailAdd{ingo.schienbein@lpsc.in2p3.fr}

\abstract{
We present the calculation of the NLO QCD corrections to the electroweak
production of top-antitop pairs at the CERN LHC in the presence of a new
neutral gauge boson. The corrections are implemented in the parton shower
Monte Carlo program POWHEG. Standard Model (SM) and new physics interference
effects are properly taken into account. QED singularities, first appearing
at this order, are consistently subtracted.
Numerical results are presented for SM and $Z'$ total cross sections and
distributions in invariant mass, transverse momentum, azimuthal angle and
rapidity of the top-quark pair. The remaining theoretical uncertainty from
scale and PDF variations is estimated, and the potential of the charge
asymmetry to distinguish between new physics models is investigated for the
Sequential SM and a leptophobic topcolor model.}


\begin{document} 
\maketitle
\flushbottom

\clearpage
\section{Introduction}
\label{sec:1}




The Standard Model (SM) of particle physics is a very successful theory describing a
wealth of experimental data up to collision energies of 13 TeV reached at CERN's Large
Hadron Collider (LHC). This includes the recent observation of a Higgs-like particle with
a mass of 125 GeV that seems to corroborate the simplest description of electroweak symmetry
breaking \cite{Aad:2012tfa,Chatrchyan:2012ufa,Aad:2015zhl}. However, the SM is based on
the unintuitive semi-simple gauge group SU(3)$_C\times$SU(2)$_L\times$U(1)$_Y$, that
together with the running behavior of the associated gauge couplings intriguingly
points towards a larger unification at some higher mass scale. The simple gauge group
SU(5) can accomodate the complete SM gauge group and its 15 fermions, but not a right-handed
neutrino, and it is in addition strongly disfavored by searches for proton decay. It also does
not allow to restore parity symmetry and does not provide a natural solution to the
neutrino mass hierarchy. Both of these important and perhaps related problems are solved
in simple gauge groups of higher rank like $E_6$ or SO(10), that can be broken consecutively
as in $E_6\to$SO(10)$\times$U(1)$_\psi$ and SO(10)$\to$SU(5)$\times$U(1)$_\chi$, respectively.
Parity restoration is achieved in left-right symmetric models, SU(3)$_C\times$SU(2)$_L\times$%
SU(2)$_R\times$U(1)$_Y$, which together with other models of similar group structure, but
different quantum number assignments form a class of general lower-scale models, commonly called
$G(221)$ models. They have recently been classified \cite{Hsieh:2010zr}, and their
phenomenology has been studied not only at the LHC \cite{Jezo:2012rm,Jezo:2014wra,Jezo:2015rha},
but also
in ultrahigh-energy cosmic rays \cite{Jezo:2014kla}. Common to all these possible extensions
of the SM is their prediction of a new heavy neutral gauge boson ($Z'$), that is associated with
the additional SU(2) or U(1) subgroup after symmetry breaking \cite{Langacker:2008yv,%
Agashe:2014kda}. In many cases, the $Z'$ boson can decay leptonically, making it a prime object of
experimental searches at the LHC. For simplification, these searches are mostly based on
the (theoretically unmotivated) Sequential SM (SSM), where the $Z'$ boson couples to other SM
particles like the SM $Z$ boson. In this model and the leptonic (i.e.\ Drell-Yan) channel, the
ATLAS and CMS collaborations have already excluded $Z'$ bosons with masses below
2.90 TeV \cite{Aad:2014cka} and 2.96 TeV \cite{CMS:2013qca}, respectively. For a recent overview
of experimental mass limits see Ref.\ \cite{Jezo:2014wra}, where it is also shown that
for certain $G(221)$ models the mass limits are enhanced to 3.2-4.0 TeV, when higher-order
QCD corrections are included.


In this paper, we focus not only on the SSM, but also on a
situation where the $Z'$ boson does not couple to leptons,
but preferentially to top quarks, so that the above mass limits are invalidated.
Models of the $G(221)$ class, where processes of the Drell-Yan type are inaccessible
at the LHC, include leptophobic (LP), hadrophobic (HP) and fermiophobic (FP) models,
whereas left-right (LR), un-unified (UU) and non-universal (NU) models remain accessible.
The LP model with a $W'$-boson mass of about 2 TeV has been put forward as a possible
explanation for the excesses of $WZ$ and $Wh$ production observed recently by ATLAS and
CMS at the LHC \cite{Gao:2015irw}.
As the heaviest particle in the SM with a mass of 173 GeV \cite{ATLAS:2014wva}, the top quark 
may very well play a special role in electroweak symmetry breaking. This motivates, e.g.,
the NU model, where the first and second SU(2) gauge groups couple exclusively to the
first/second and third generation fermions, respectively.
It also motivates models with new strong dynamics such as the topcolor
model \cite{Hill:1991at,Hill:1994hp}, which can generate a large top-quark mass through the
formation of a top-quark condensate. This is achieved by introducing
a second strong SU(3) gauge group which couples preferentially to the
third generation, while the
original SU(3) gauge group couples only to the first and second generations. To block the
formation of a bottom-quark condensate, a new U(1) gauge group and associated $Z'$ boson are
introduced. Different couplings of the $Z'$ boson to the three fermion generations then
define different variants of the model \cite{Harris:1999ya}. A popular choice with the LHC
collaborations is the leptophobic topcolor model (also called Model IV in the reference cited
above) \cite{Harris:2011ez}, where the $Z'$ couples only to the first and third generations of
quarks and has no significant couplings to leptons, but an experimentally accessible cross
section.


The strongest limits on $Z'$ bosons arise of course from their Drell-Yan like decays
into electrons and muons at the LHC. This is due to the easily identifiable experimental
signatures
\cite{Jezo:2014wra}. The top-pair signature is more difficult, as top quarks decay to $W$ bosons
and bottom quarks, where the latter must be tagged and the two $W$ bosons may decay hadronically,
i.e.\ to jets, or leptonically, i.e.\ into electrons or muons and missing energy carried away by
a neutrino. In addition and in contrast to the Drell-Yan process, the electroweak top-pair
production cross section obtains QCD corrections not only in the initial, but also in the
final state. For conclusive analyses,
precision calculations are therefore extremely important
to reduce theoretical uncertainties, arising from variations of the renormalization and
factorization scales $\mu_r$ and $\mu_f$ and of the parton density functions (PDFs)
$f_{a/p}(x_a,\mu_f)$, and for an accurate description of the possible experimental signal and the
SM backgrounds.

At the LHC, the hadronic top-pair production cross section
\bea
 \sigma&=& \sum_{ab}\int f_{a/p}(x_a,\mu_f)f_{b/p}(x_b,\mu_f)\,{d\sigma_{ab}\over d t}(\mu_r)\,dt\, dx_a dx_b
\eea
obtains up to next-to-leading order (NLO) the contributions
\bea
 \sigma_{ab}(\mu_r) &=& \sigma_{2;0}(\alpha_{S}^2)
 + {\color{red} \sigma_{0;2}(\alpha^2)}
 + \sigma_{3;0}(\alpha_S^3)
 + \sigma_{2;1}(\alpha_S^2\alpha)
 + {\color{red}\sigma_{1;2}(\alpha_{S}\alpha^2)}
 + \sigma_{0;3}(\alpha^3)\,,
 \label{eq:1.2}
\eea
where the numerical indices represent the powers of the strong coupling $\alpha_S(\mu_r)$ and of
the electromagnetic coupling $\alpha$, respectively.
The first and third terms representing the SM QCD background
processes $q\bar{q},gg\to t\bar{t}$ and their NLO QCD corrections, including the
$qg$ channel, have been computed in the late 1980
\cite{Nason:1987xz,Nason:1989zy,Beenakker:1988bq,Beenakker:1990maa}.
Furthermore, NLO predictions for heavy quark correlations have
been presented in \cite{Mangano:1991jk}, and the spin correlations 
between the top quark and antiquark have been studied in the early
2000s \cite{Bernreuther:2001rq,Bernreuther:2004jv}.
The fourth term represents the electroweak
corrections to the QCD backgrounds, for which a gauge-invariant subset was first investigated
neglecting the interferences between QCD and electroweak interactions arising from box-diagram
topologies and pure photonic contributions \cite{Beenakker:1993yr} and later including also
additional Higgs boson contributions arising in 2-Higgs doublet models (2HDMs) \cite{Kao:1999kj}.
The rest of the electroweak corrections was calculated in a subsequent series of papers and
included also $Z$-gluon interference effects and QED corrections with real and virtual photons
\cite{Kuhn:2005it,Moretti:2006nf,Bernreuther:2005is,Bernreuther:2006vg,Hollik:2007sw}. In this
paper, we focus on
the second and fifth terms in Eq.\ (\ref{eq:1.2}) (highlighted in red), i.e.\ the contribution
$\sigma_{0;2}$ for the $Z'$ signal and its interferences with the photon and SM $Z$ boson and the
corresponding QCD corrections $\sigma_{1;2}$. Due to the resonance of the $Z'$ boson, we expect
these terms to be the most relevant for new physics searches. A particular advantage of this
choice is that the calculation of $\sigma_{1;2}$ can then be carried out in a model-independent
way as long as the $Z'$ couplings are kept general, whereas the fourth term $\sigma_{2;1}$ is
highly model-dependent due to the rich structure of the scalar sector in many models. The sixth
term in Eq.\ (\ref{eq:1.2}) is suppressed by a relative factor $\alpha/\alpha_s$ with respect
to the fifth and thus small.

The production of $Z'$ bosons (and Kaluza-Klein gravitons) decaying to top pairs has been
computed previously in NLO
QCD by Gao et al.\ in a factorized approach, i.e.\ neglecting all SM interferences and quark-gluon
initiated diagrams with the $Z'$ boson in the $t$-channel, and for purely vector- and/or
axial-vector-like couplings as those of the SSM \cite{Gao:2010bb}. We have verified that we
can reproduce their $K$-factors (i.e.\ the ratio of NLO over LO predictions) of 1.2 to 1.4
(depending on the $Z'$ mass) up to 2\%, if we reduce our calculation to their theoretical
set-up and employ their input parameters. Their result has triggered the Tevatron and LHC
collaborations to routinely use a $K$-factor of 1.3 in their experimental analyses (see below).
The factorized calculation by Gao et al.\ has been confirmed previously in an independent NLO
QCD calculation by Caola et al.\ \cite{Caola:2012rs}. Like us, these last authors include also the
additional quark-gluon initiated processes and show that after kinematic cuts
they reduce the $K$-factor by about 5
\%. However, they still do not include the additional SM interferences, which they claim to be
small for large $Z'$-boson masses. As we will show, this is not always true due to logarithmically
enhanced QED contributions from initial photons. In contrast to us, they also include
top-quark decays in the narrow-width approximation with spin correlations and
box-diagram
corrections to interferences of the electroweak and QCD Born processes ($\sigma_{2;1}$ in Eq.\
(\ref{eq:1.2})), which are, however, only relevant for very broad resonances.
If the (factorizable) QCD corrections to the top-quark decay are included,
the $K$-factor is reduced by an additional 15\%. The globally smaller $K$-factor of Caola
et al.\ is thus explained by calculational aspects and not by different choices of input
parameters.

The SM backgrounds are today routinely calculated not just in NLO QCD, but at NLO combined with
parton showers (PS), e.g.\ within the framework of MC@NLO or POWHEG \cite{Frixione:2002ik,Frixione:2007vw}.
A particularly useful tool is the POWHEG BOX, in which new processes can be implemented once
the spin- and color-correlated Born amplitudes along with their virtual and real NLO QCD
corrections are known and where the regions of singular radiation are then automatically
determined \cite{Alioli:2010xd}. Calculations of this type have already been performed by us
in the past for the Drell-Yan like production of $Z'$ bosons \cite{Fuks:2007gk}, heavy-quark
production in the ALICE experiment \cite{Klasen:2014dba}, and the 
associated production of top quarks and charged Higgs bosons \cite{Weydert:2009vr,Klasen:2012wq}.
In this work, we provide a calculation of the $Z'$ signal with a final top-quark pair at the
same level of accuracy, including all interferences with SM $Z$ bosons and photons as well as
the logarithmically enhanced QED contributions from initial-state photons, which we will
discuss in some detail. We also present details about the spin- and color-correlated Born
amplitudes, the treatment of $\gamma_5$ and renormalization procedure in our calculation of
the virtual corrections, as well as the validation of our NLO+PS calculation, which we have
performed with the calculation for $Z'$ bosons of Gao et al.\ at NLO \cite{Gao:2010bb}
and for tree-level and one-loop SM matrix elements with MadGraph5\_aMC@NLO \cite{Alwall:2014hca}
and GoSam \cite{Cullen:2011ac}.


Experimental searches for resonant top-antitop production have been performed at the Tevatron
and at the LHC mostly for the leptophobic topcolor model with a $Z'$-boson coupling only
to first and third generation quarks \cite{Harris:1999ya,Harris:2011ez}. In this model,
the LO cross section is controlled by three parameters: the ratio of the two U(1) coupling
constants, $\cot\theta_H$, which should be large to enhance the condensation of top quarks, but
not bottom quarks, and which also controls both the $Z'$ production cross section and decay
width, as well as the relative
strengths $f_1$ and $f_2$ of the couplings of right-handed up- and down-type quarks with respect
to those of the left-handed quarks. The LO cross sections for this model are usually computed for
a fixed small $Z'$ width, $\Gamma_{Z'}=1.2\%\times m_{Z'}$, effectively setting the parameter
$\cot\theta_H$, and the choices $f_1=1$, $f_2=0$, which maximize the fraction of $Z'$ bosons that
decay into top-quark pairs.
We have verified that we can reproduce the LO numerical results in the paper by Harris and Jain
\cite{Harris:2011ez} for $Z'$ masses above 1 TeV and relative widths of 1\% and 1.2\%,
but not 10\%, if we neglect all SM interferences. As stated above, the LO cross sections are
routinely multiplied by the experimental collaborations by a $K$-factor of 1.3 \cite{Gao:2015irw}.
At the Tevatron with center-of-mass energy $\sqrt{S}=1.96$ TeV and in the lepton+jets
top-quark decay channel, CDF and D0 exclude $Z'$ bosons with masses up to 0.915 TeV
\cite{Aaltonen:2012af} and 0.835 TeV \cite{Abazov:2011gv}, respectively. The weaker
D0 limit can be
explained by the fact that CDF use the full integrated luminosity of 9.45 fb$^{-1}$, while
D0 analyze only 5.3 fb$^{-1}$ and furthermore do not use a $K$-factor for the signal cross
section.
At the LHC, the ATLAS and CMS collaborations have analyzed 20.3 fb$^{-1}$ and 19.7 fb$^{-1}$
of integrated luminosity of the $\sqrt{S}=8$ TeV LHC run employing the $K$-factor of 1.3.
The result is that narrow leptophobic
topcolor $Z'$ bosons are excluded below masses of 1.8 TeV and 2.4 TeV, respectively
\cite{Aad:2015fna,Khachatryan:2015sma}. At the LHC, the CMS limit is currently
considerably stronger than the one by ATLAS despite the slightly smaller exploited luminosity.
The reason is that CMS performed a combined analysis of all top-quark decay
channels (dilepton, lepton+jets and all hadronic), while ATLAS analyzed only the
lepton+jets channel. For $\Gamma_{Z'}=10\%\times m_{Z'}$, the CMS mass limit is even stronger and
is found to be 2.9 TeV.
%
%
%
We emphasize that the narrow width assumption employed in most experimental analyses
need not be realized in nature and that in this case a proper treatment of SM interference
terms as provided in our full calculation is required.

The LHC has just resumed running with an increased center-of-mass energy of 13 TeV,
which is planned to be increased to 14 TeV in the near future. We therefore provide
numerical predictions in this paper for both of these energies and for two benchmark
models, i.e.\ the SSM and the leptophobic topcolor model. The predictions for the SSM
are readily obtained by taking over the $Z'$-boson couplings from the SM, with the consequence
of again a relatively small width $\Gamma_{Z'}\simeq 3\% \times m_{Z'}$ for $Z'$ masses
between 3 and 6 TeV. We focus on the invariant-mass distribution of the top-quark pair,
which is the main observable exploited
for resonance (and in particular $Z'$-boson) searches, but also show results for the
distributions that are most sensitive to soft parton radiation beyond NLO, i.e.\
the transverse momentum $p_{t\bar{t}}$ of the top-antitop pair and their relative azimuthal angle
$\phi_{t\bar{t}}$. The forward-backward asymmetry $A_{FB}$ of top-antitop events with positive
vs.\ negative rapidity difference between the two has also been suggested as
a very useful observable to distinguish among different models \cite{Kamenik:2011wt}.
At the Tevatron (a
$p\bar{p}$ collider, where top quarks are produced predominantly in the direction of the
proton beam), long-standing discrepancies of CDF and D0 measurements with the SM prediction
at NLO \cite{Aaltonen:2011kc,Abazov:2011rq} have triggered numerous suggestions of new
physics contributions \cite{Kamenik:2011wt}, e.g.\ of light $Z'$ bosons coupling in a flavor
non-diagonal way to up and top quarks \cite{Buckley:2011vc}. Only recently the SM
prediction at next-to-next-to-leading order (NNLO) \cite{Czakon:2014xsa} has been
brought in agreement with the newest inclusive measurement by CDF \cite{Aaltonen:2012it} and 
differential measurement by D0 \cite{Abazov:2014cca}. At the LHC (a $pp$ collider), a
charge asymmetry $A_C$ can be defined with respect to the difference in absolute value of
the top and antitop rapidities \cite{AguilarSaavedra:2012rx}. We therefore also provide
numerical predictions for this observable in our two benchmark models and at current and
future LHC center-of-mass energies.


Our paper is organized as follows: In Sec.\ \ref{sec:2} we present analytical results
of our calculations at LO and the NLO virtual and real corrections, including details
about SM interference terms, our treatment of $\gamma_5$, our renormalization procedure
and the subtraction method employed for the soft and collinear divergences in the
real corrections. In Sec.\ \ref{sec:3} we discuss the implementation of our calculation
in POWHEG and present in particular the color- and spin-correlated Born
amplitudes, the definition of the finite remainder of the virtual corrections,
the implementation of the real corrections with a focus on the rather involved
treatment of QED divergences, and the 
validation of our tree-level matrix elements in the SM against those of the automated
tool MadGraph5\_aMC@NLO \cite{Alwall:2014hca} 
and of the virtual corrections against those of GoSam \cite{Cullen:2011ac}
as well
as of our numerical pure $Z'$-boson results against those obtained
by Gao et al.\ and Caola et al. Our new numerical predictions for the LHC are shown and
discussed in Sec.\ \ref{sec:4}, and Sec.\ \ref{sec:5} contains our conclusions.
Several technical details of our calculation can be found in the Appendix.

\section{NLO QCD corrections to electroweak top-pair production}
\label{sec:2}

In this section, we present in detail our calculation of the NLO QCD corrections
to electroweak top-pair production through photons, SM $Z$ bosons and additional
$Z'$ bosons with generic vector and axial-vector couplings to the SM fermions.
We generate all Feynman diagrams automatically with QGRAF \cite{Nogueira:1991ex}
and translate them into amplitudes using DIANA \cite{Tentyukov:1999is}. The traces
of the summed and squared amplitudes with all interferences are then calculated
in the Feynman gauge and $D=4-2\varepsilon$ dimensions in order to regularize the
ultraviolet (UV) and infrared (IR) divergences using FORM \cite{Vermaseren:2000nd}.
Traces involving the Dirac matrix $\gamma_5$ are treated in the Larin
prescription \cite{Larin:1993tq} by replacing $\gamma_{\mu}\gamma_5 = i\frac{1}{3!}
\varepsilon_{\mu\nu\rho\sigma}\gamma^{\nu}\gamma^\rho\gamma^\sigma$.
To restore the Ward identities and thus preserve gauge invariance at one loop, we
perform an additional finite renormalization for vertices involving $\gamma_{5}$.

\subsection{Leading-order contributions}
\label{sec:2.1}

\begin{figure}[!h]
 \centering
 \includegraphics[width=0.25\textwidth]{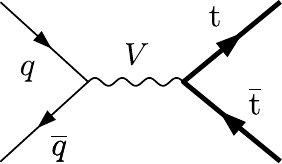}
 \caption{Tree-level Feynman diagrams of order $\gr{O}(\alpha)$ contributing to electroweak
 top-pair production through vector bosons $V$, i.e.\ photons ($\gamma$), SM $Z$ bosons and
 new $Z'$ bosons.}
 \label{fig:01}
\end{figure}

The leading-order (LO) Feynman diagrams contributing to the electroweak production
of top-quark pairs at ${\cal O}(\alpha)$ through photons, SM  $Z$ bosons and
new $Z'$ bosons are shown summarily in Fig.\ \ref{fig:01}.
The cross section $d\sigma/dt$, differential in the Mandelstam variable $t$ denoting the
squared momentum transfer, is then obtained by summing all three corresponding amplitudes,
squaring them, summing/averaging them over final-/initial-state spins and colors and multiplying
them with the flux factor $1/(2s)$ of the incoming and the differential phase space $1/(8\pi s)$
of the outgoing particles. The Mandelstam variable $s$ denotes the squared partonic
center-of-mass energy. The result, given here for brevity only in four and not $D$ dimensions,
is
\bea
 {d\sigma_{q\bar{q}}\over dt}&=& \frac{1}{2s} \frac{1}{8\pi s} {B}_{q\bar{q}} \\
 &=& \frac{1}{2s} \frac{1}{8\pi s} \sum_{V,V'}
 \frac{2e^4 D_V D_{V^\prime}}{s_W^4} \left\{s(t-u)
 \left( A_{V}^{\Pq} B_{V^\prime}^{\Pq} + A_{V^\prime}^{\Pq} B_{V}^{\Pq} \right)
 \left( A_{V}^{\Pqt} B_{V^\prime}^{\Pqt} + A_{V^\prime}^{\Pqt} B_{V}^{\Pqt} \right)
 \right.\nonumber\\
 &+& \left. \left( A_{V}^{\Pq} A_{V^\prime}^{\Pq}\! +\! B_{V}^{\Pq} B_{V^\prime}^{\Pq} \right)
 \left[ \left( t^2 + u^2 + 4sm_{\Pqt}^2 - 2m_{\Pqt}^4 \right)
 A_{V}^{\Pqt} A_{V^\prime}^{\Pqt} + \left( t^2 + u^2 - 2m_{\Pqt}^4 \right)
 B_{V}^{\Pqt} B_{V^\prime}^{\Pqt} \right] \right\}\nonumber\\
 &\times& \left\{ \left[ (s\! -\! m_{V}^2) (s\! -\! m_{V^\prime}^2) + m_{V} m_{V^\prime}
 \Gamma_{V} \Gamma_{V^\prime} \right] + i \left[ (s\! -\! m_{V}^2) m_{V^\prime}
 \Gamma_{V^\prime} - (s\! -\! m_{V^\prime}^2) m_{V} \Gamma_{V} \right] \right\},\nonumber
 \label{eq:BornRes1}
\eea
where ${B}_{q\bar{q}}$ is the modulus squared of the Born amplitude averaged/summed over
initial/final spins and colors, $V,V^\prime \in\{\Pphoton,\PZ,\PZprime\}$, the superscript
$q$ denotes the flavor of the incoming massless quarks, $s,\ t,\ u$ are the partonic
Mandelstam variables, and $m_t$ is the top-quark mass. Note that we
use the Pauli metric, in which the dot-product has an overall minus sign with respect to
the Bjorken-Drell metric \cite{Veltman:1994wz}. The terms $D_{V},D_{V^\prime}$ stem from the
propagator denominators and take the usual form 
\begin{equation}
 D_{\Pphoton}  = \frac{1}{s^2},
 \ D_{\PZ}  = \frac{1}{(s-m_{\PZ}^2)^2+m_{\PZ}^2\Gamma_{\PZ}^2},
 \ D_{\PZprime} = \frac{1}{(s-m_{\PZprime}^2)^2+m_{\PZprime}^2\Gamma_{\PZprime}^2}\,.
\label{eq:BornRes2}
\end{equation}
To take into account the finite widths of the $Z$ and $Z^\prime$ bosons, we have introduced
complex masses $m_{Z(Z^{\prime})}\rightarrow m_{Z(Z^{\prime})} - i \Gamma_{Z(Z^\prime)}/2$
with the consequence that $m_{Z(Z^{\prime})}^{2} \to m_{Z(Z^\prime)}^2 - \Gamma^{2}_{Z(Z^\prime)}/4$.
The coefficients $A^{\Pq}_{V(V^\prime)},B^{\Pq}_{V(V^\prime)},A^{\Pqt}_{V(V^\prime)},B^{\Pqt}_{V(V^\prime)}$
are proportional to the axial ($A$) and vector ($B$) couplings of the various gauge bosons
to the massless quarks ($\Pq=u,d,s,c,b$) and the top quark ($\Pqt$),
\begin{align}
	 A_{\Pphoton}^q & = s_W Q_q, & A_{\Pphoton}^{\Pqt} & = s_W Q_t, & B_{\Pphoton}^{q} & = 0, & B_{\Pphoton}^{\Pqt} & = 0, \nonumber \\
	 A_{\PZ}^q & = \frac{a_{\PZ}^q}{4c_W }, & A_{\PZ}^{\Pqt} & = \frac{a_{\PZ}^t}{4c_W}, & B_{\PZ}^{q} & = \frac{b_{\PZ}^{q}}{4c_W}, & B_{\PZ}^{\Pqt} & = \frac{b_{\PZ}^{t}}{4c_W}, \nonumber \\
	 A_{\PZprime}^q & = \frac{a_{\PZprime}^{q}}{4c_W}, & A_{\PZprime}^{\Pqt} & = \frac{a_{\PZprime}^{\Pqt}}{4c_W}, & B_{\PZprime}^{q} & = \frac{b_{\PZprime}^{q}}{4c_W}, & B_{\PZprime}^{\Pqt} & = \frac{b_{\PZprime}^{\Pqt}}{4c_W}\,, 
\end{align}
where $s_W\,(c_W)$ are the sine (cosine) of the weak mixing angle $\theta_W$, $Q_q$ is
the fractional
charge of quark flavor $q$, and $a_V^q$ and $b_V^q$ are the model-dependent vector and axial-vector
couplings of the $Z$ and $Z'$ bosons, e.g. $a_Z^u=1-8/3s_W^2$, $a_Z^d=4/3s_W^2-1$, $b_Z^u=1$,
$b_Z^d=-1$ for all up- and down-type quarks in the SM. Although individual interference terms
may contain imaginary parts, they cancel as expected after summation.

\subsection{One-loop virtual corrections}
\label{sec:2.2}

\begin{figure}[!h]
 \centering
 \includegraphics[width=0.5\textwidth]{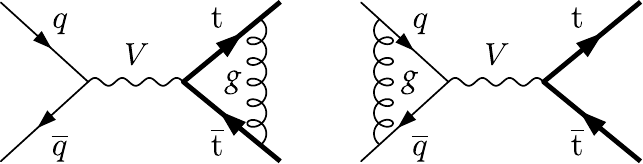}
 \caption{One-loop Feynman diagrams of order $\gr{O}(\alpha_{S}\alpha)$ contributing to
 electroweak top-pair production.}
 \label{fig:02}
\end{figure}

\begin{figure}[!h]
 \centering
 \includegraphics[width=0.25\textwidth]{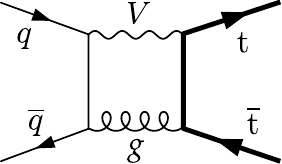}
 \caption{Example of a box diagram of $\gr{O}(\alpha_S\alpha)$ leading to a vanishing
 contribution. This diagram would, however, contribute to electroweak corrections to
 the QCD Born processes.}
 \label{fig:03}
\end{figure}

The one-loop virtual corrections contributing to electroweak top-pair production
at $\gr{O}( \alpha_s\alpha^2)$ originate from the interferences among the one-loop
diagrams shown in Fig.~\ref{fig:02} with the tree-level diagrams in Fig.~\ref{fig:01}.
Note that one-loop electroweak corrections to the QCD process $q\bar q \to g^*
\to t \bar{t}$ have zero interference with the electroweak diagrams in Fig.~\ref{fig:01}, since
such contributions are proportional to the vanishing color trace $\mathrm{Tr}(T^a)$.
In particular, the interference term of the box diagram in Fig.~\ref{fig:03} with the amplitudes
in Fig.~\ref{fig:01} vanishes, whereas it would of course contribute at $\gr{O}(\alpha_s^2
\alpha)$.

As already mentioned, the virtual amplitudes are regularized dimensionally. The appearing 30
distinct loop integrals are then reduced to a basis of three master integrals
using integration-by parts identities \cite{Tkachov:1981wb,Chetyrkin:1981qh} in the form of the
Laporta algorithm \cite{Laporta:2001dd} as implemented in the public tool REDUZE
\cite{2010CoPhC.181.1293S,vonManteuffel:2012np}.
One is thus left with the evaluation of three master integrals: the
massive tadpole, the equal-masses two-point function, and the massless 
two-point function. The solutions of these integrals are well known \cite{Hooft:1978xw}.
For completeness, we provide their analytic expressions in App.\ \ref{sec:a}.

In dimensional regularization, the UV and IR singularities in the virtual corrections appear
as poles of $1/\varepsilon$ and $1/\varepsilon^2$. Since neither the couplings nor the top-quark
mass have to be renormalized at NLO, the UV singularities can be removed by simply adding the Born
cross section multiplied with the quark wave-function renormalization constants
\begin{equation}
 \sum_{\psi\in\{\Pq,\bar \Pq,\Pqt,\bar \Pqt\}}\frac{1}{2}\delta Z_\psi \,.
\end{equation}
We use the on-shell renormalization scheme, in which $\delta Z_q=0$ for the
initial-state massless quarks and
\beq
 \delta Z_t=(4\pi)^\varepsilon\Gamma(1+\varepsilon)
 \left({\mu_r^2\over m_t^2}\right)^\varepsilon {C_F\alpha_s\over\pi}\left(-{3\over4\varepsilon}
 -{1\over1-2\varepsilon}\right)
\eeq
for the final-state top quarks.
Since we are using the Larin prescription for $\gamma_5$ (see above), we must perform
an additional finite renormalization to restore the Ward identities. The corresponding
constant has been calculated up to three loops in the $\MSb$ scheme \cite{Larin:1993tq}.
At one loop, it reads
\beq
 \delta Z_5=-{C_F\alpha_s\over\pi}
\eeq
and multiplies all appearing factors of $\gamma_5$.
Once the UV divergences are renormalized, we are left with infrared collinear and
soft divergences that match the correct structure given for instance in Refs.\
\cite{Catani:2002hc,Frixione:1995ms}. For completeness, we provide the analytic
expressions of the IR poles in App.\ \ref{sec:b}.

\subsection{Real emission corrections}
\label{sec:2.3}

\begin{figure}[!h]
 \centering
 \includegraphics[width=\textwidth]{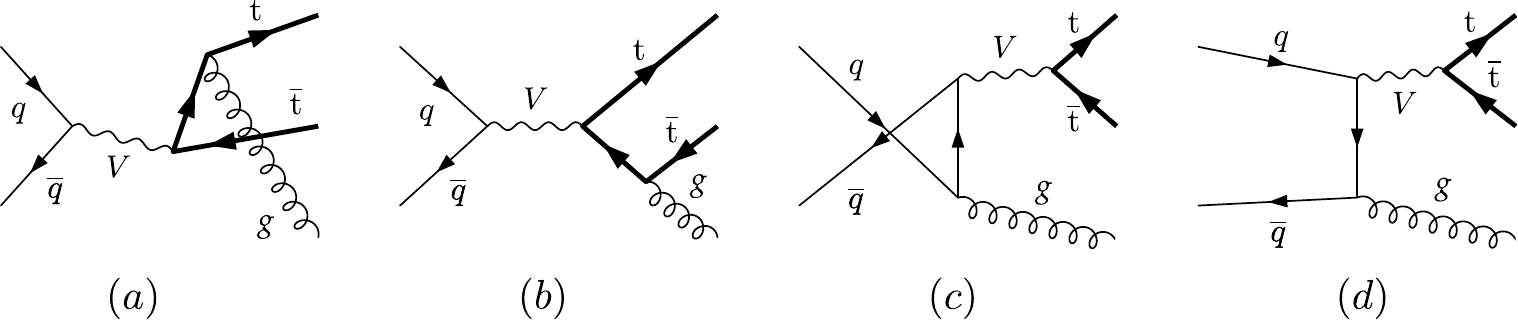}
 \caption{Diagrams contributing to the $q + \bar q \to t + \bar t + g$ subprocess at
 $\gr{O}(\alpha_{S}\alpha^2)$ with $V\in \{\gamma,Z,Z^{\prime}\}$.}
 \label{fig:04}
\end{figure}

\begin{figure}[!h]
 \centering
 \includegraphics[width=\textwidth]{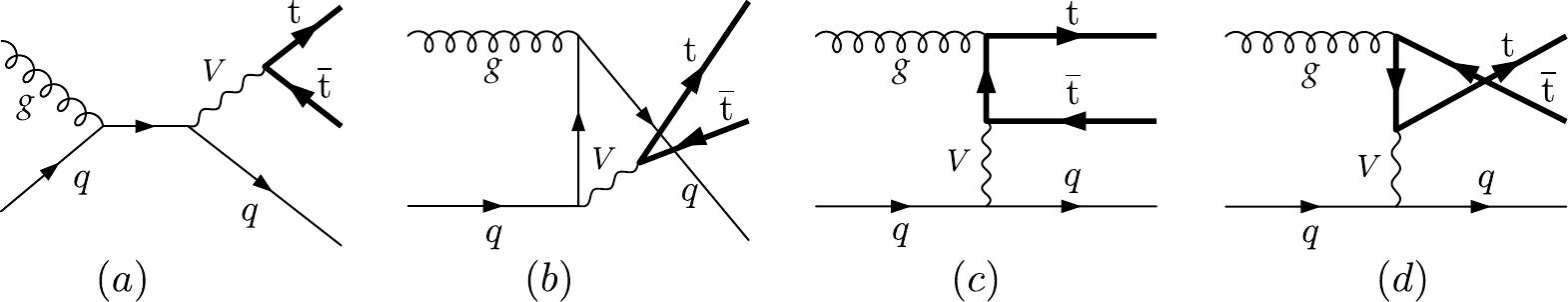}
 \caption{Diagrams contributing to the $g+q \to t + \bar t + q$ subprocess at $\gr{O}(\alpha_{S}
 \alpha^2)$ with $V\in \{\gamma,Z,Z^{\prime}\}$. Similar diagrams contribute to the $g \bar q$
 channel.}
 \label{fig:05}
\end{figure}

At $\gr{O}(\alpha_S\alpha^2)$, the following $2\rightarrow 3$ tree-level processes 
contribute:
\begin{inparaenum}[(i)]
\item $q +\bar q \to t + \bar t + g$ and
\item $g + q(\bar q) \to t + \bar t + q(\bar q)$.
\end{inparaenum}
The corresponding Feynman diagrams are depicted in  Figs.~\ref{fig:04} and \ref{fig:05}.
In the $q \bar q$ channel, the diagrams in Figs.~\ref{fig:04} (a) and (b) only have a singularity
when the gluon emitted from the heavy top-quark line becomes soft, whereas those in Figs.\
\ref{fig:04} (c) and (d) diverge when the radiated gluon becomes soft and/or collinear to the
emitting light quark or antiquark. The $g q$ and $g\bar q$ channels exhibit at most collinear
singularities. While the diagram in Fig.\ \ref{fig:05} (a) is completely finite, the outgoing
quarks in Figs.~\ref{fig:05} (b) or (c) and (d) can become collinear to the initial gluon or
quark.

\begin{figure}[!h]
 \centering
 \includegraphics[width=0.5\textwidth]{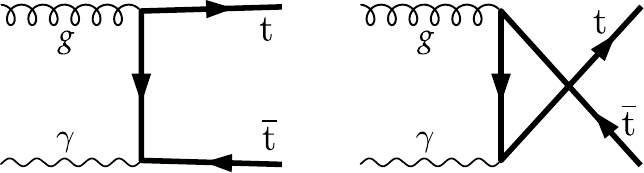}
 \caption{Photon-induced top-pair production of $\gr{O}(\alpha_S\alpha)$. These
 diagrams must be added for a consistent subtraction of the collinear singularities.}
 \label{fig:06}
\end{figure}

As a consequence of the KLN theorem, the soft and soft-collinear divergences
cancel in the sum of the real and virtual cross sections, while the collinear
singularities are absorbed into the parton distribution functions (PDFs) by
means of the mass factorization procedure. The singularities in the real
corrections are removed in the numerical phase space integration by subtracting
the corresponding unintegrated counter terms \cite{Catani:2002hc,Frixione:1995ms}.
The fact that the collinear divergences appearing in Figs.~\ref{fig:05} (c) and (d) involve a
photon propagator has two consequences:
\begin{inparaenum}[(i)]\item we have to introduce a PDF for the photon inside the proton and
\item the corresponding underlying Born process shown in Fig.~\ref{fig:06}, $g+\gamma \to
t + \bar t$, must be included in the calculation.
\end{inparaenum}
The squared modulus of the corresponding Born amplitude, averaged/summed over initial/final
state spins and colors, is
\bea
 {B}_{g\gamma}&=&
 16\pi^2\alpha_s\alpha Q_t^2 
 \left[{t_t\over u_t}+{u_t\over t_t}+{4m_t^2s\over t_tu_t}
 \left(1-{m_t^2 s\over t_tu_t}\right)\right],
\eea
with $Q_t$ the fractional electric charge of the top quark (2/3),
$N_C=3$, $C_F=4/3$, $t_t=t-m_t^2$ and $u_t=u-m_t^2$. Although this process is formally of 
$\gr{O}(\alpha_S\alpha)$ and thus contributes to $\sigma_{1;1}$, it is multiplied by a
photon distribution inside the proton of $\gr{O}(\alpha)$, so that the hadronic subprocess
$p+p\to g+\gamma \to t + \bar t$ is effectively of $\gr{O}(\alpha_S\alpha^2)$.
As we will see in Sec.~\ref{sec:4}, this channel is indeed 
numerically important. 

\section{POWHEG implementation}
\label{sec:3}

We now turn to the implementation of our NLO corrections to electroweak top-pair
production, described in the previous section, in the NLO+PS program POWHEG
\cite{Alioli:2010xd}. We thus combine the NLO precision of our analytical
calculation with the flexibility of parton shower Monte Carlo programs like PYTHIA
\cite{Sjostrand:2007gs} or HERWIG \cite{Corcella:2000bw} that are indispensible tools
to describe complex multi-parton final states, their hadronization, and particle
decays at the LHC. Since the leading emission is generated both at NLO and with the
PS, the overlap must be subtracted, which is achieved using the POWHEG method
\cite{Frixione:2007vw} implemented in the POWHEG BOX \cite{Alioli:2010xd}.
In the following, we describe the required color- and
spin-correlated Born amplitudes, the definition and implementation of the finite
remainder of the virtual corrections, and the real corrections with a focus on
the subtleties associated with the encountered QED divergences. All other aspects
such as lists of the flavor structure of the Born and real-emission processes, the
Born phase space, and the four-dimensional real-emission squared matrix elements
have either already been discussed above or are trivial to obtain following the POWHEG
instructions \cite{Alioli:2010xd}. We end this section with a description of the
numerical validation of our implementation.

\subsection{Color-correlated Born amplitudes}

The automated calculation of the subtraction terms in POWHEG requires the knowledge
of the color correlations between all pairs of external legs $i,j$. The color-correlated
squared Born amplitude $\gr{B}_{ij}$ is formally defined by
\bea
 \gr{B}_{ij} &=&-N
 \sum_{\begin{array}{c}\text{\scriptsize spins}\\[-0.18cm]\text{\scriptsize colors}\end{array}}
 \gr{M}_{\{c_k\}}\left( \gr{M}^\dagger_{\{c_k\}} \right)_{
 \begin{array}{c} c_i\rightarrow c^\prime_i\\c_j\rightarrow c^\prime_j\end{array}}
 T^a _{c_i,c^\prime_i}T^a_{c_j,c^\prime_j}\,,
 \label{eq:3.1}
\eea
where $N$ is the normalization factor for initial-state spin/color averages and final-state
symmetrization, $\gr{M}_{\{c_k\}}$ is the Born amplitude and $\{c_k\}$ are the color indices of all
external colored particles. The suffix of $(\gr{M}^{\dagger}_{\{c_k\}})$ indicates that the color
indices of partons $i,j$ must be replaced with primed indices. For incoming quarks and
outgoing antiquarks $T^{a}_{c_i,c_i'}=t^a_{c_ic_i'}$, where $t$ are the color
matrices in the fundamental representation of SU(3), for incoming antiquarks and
outgoing quarks $T^{a}_{c_ic_i'}=-t^a_{c_i'c_i}$, and for gluons $T^{a}_{c_ic_i'}=if_{c_i a c_i'}$,
where $f_{abc}$ are the structure constants of SU(3).
For the $q\bar{q}$-initiated electroweak top-pair production, one obtains in a
straightforward way
\beq
  \gr{B}_{ij}  = C_{F} {B}_{q\bar{q}}
\eeq
for two incoming ($i,j=q,\bar{q}$) or outgoing ($i,j=t,\bar{t}$) particles and zero otherwise.

As we have seen in Sec.\ \ref{sec:2.3}, we also have to include the gluon-photon induced pair
production process in order to treat the QED divergence occurring in the $gq$ real-emission
correction. We thus also have to calculate the color-correlated squared Born matrix element
for this process. The color structure of the corresponding Feynman diagrams, see Fig.\
\ref{fig:06}, factorizes in the amplitude, and we can thus directly calculate the
color-correlated in terms of the averaged/summed modulus squared of the Born matrix element
with color factor $\gr{C}=N_CC_F=(N_C^2-1)/2$. Applying Eq.\ (\ref{eq:3.1}) to all pairs of colored
external legs, we obtain
\bea
  \gr{B}_{13} &=& - \frac{1}{\gr{C}}t^{a}_{\alpha\beta}t^{a^\prime}_{\beta \alpha^{\prime}} T^{e}_{a,a^\prime}T^{e}_{\alpha\alpha^{\prime}}B_{g\gamma}  =-t^{a}_{\alpha\beta} t^{a^\prime}_{\beta\alpha^\prime}if_{aea^{\prime}}( -t^{e}_{\alpha^\prime \alpha})\frac{B_{g\gamma}}{\gr{C}}\\
  &=&-if_{a^\prime e a}\mathrm{Tr}(t^{a^\prime}t^e t^a)\frac{B_{g\gamma}}{\gr{C}} =\frac{1}{2}N_C \mathrm{Tr}(t^a t^a) \frac{B_{g\gamma}}{\gr{C}}\nonumber\\
  &=& \frac{1}{2}N_CB_{g\gamma}\,,\nonumber\\
 \gr{B}_{14} &=& \gr{B}_{13}= \frac{1}{2}N_CB_{g\gamma}\,,\\
 \gr{B}_{34} &=& \gr{B}_{43} ~=~
 -\frac{1}{\gr{C}}B_{g\gamma}t^{a}_{\alpha\beta}t^{b}_{\beta^{\prime}\alpha^{\prime}}T^{e}_{\beta\beta^\prime}T^e_{\alpha\alpha^\prime}\delta^{ab}=\mathrm{Tr}(t^at^et^at^e) \frac{1}{\gr{C}}B_{g\gamma} = \frac{-1}{2N_C}B_{g\gamma}\,.
\eea
As is easily verified, a completeness relation coming from color conservation holds:
\begin{eqnarray}
  \gr{B}_{13} + \gr{B}_{14} &=&  \left( \frac{1}{2}N_C + \frac{1}{2}N_C \right)B_{g\gamma} = N_C B_{g\gamma}\,,\nonumber\\
  \gr{B}_{34} + \gr{B}_{31} &=&  \left( \frac{-1}{2N_C} + \frac{1}{2}N_C \right)B_{g\gamma} = \frac{N_C^2-1}{2N_C} B_{g\gamma} = C_F B_{g\gamma}\,,
\end{eqnarray}
and similarly for $\gr{B}_{41}+\gr{B}_{43}$. These cross checks are also performed automatically
in POWHEG.

\subsection{Spin-correlated Born amplitudes}

The spin-correlated squared Born amplitude $B^{\mu\nu}_{j}$ only differs from zero, if leg $j$
is a gluon. It is obtained by leaving uncontracted the polarization indices of this leg, i.e.\
\beq
  \gr{B}^{\mu\nu}_{j} =N\sum_{\{i\},s_j,s^\prime_j}\gr{M}(\{i\},s_j)\gr{M}^{\dagger}(\{i\},s^{\prime}_j)(\varepsilon^{\mu}_{s_j})^{\ast}\varepsilon^{\nu}_{s^\prime_j}\,,
\eeq
where $\gr{M}(\{i\},s_j)$ is the Born amplitude, $\{i\}$ represents collectively all remaining
spins and colors of the incoming and outgoing particles, and $s_j$ is the spin of particle $j$.
The polarization vectors $\varepsilon^{\mu}_{s_j}$ are normalized according to
\begin{equation}
  \sum_{\mu,\nu} g_{\mu\nu}(\varepsilon^{\mu}_{s_j})^{\ast}\varepsilon^{\nu}_{s^\prime_j}=-\delta_{s_js^\prime_j}\,.
\end{equation}
Similarly to the color-correlated Born amplitudes, we have a closure relation, namely
\begin{equation}
  \sum_{\mu,\nu}g_{\mu\nu} \gr{B}^{\mu\nu}_j =-{B}\,,
 \label{eq:3.10}
\end{equation}
where $B$ is the squared Born amplitude after summing over all polarizations.
Since processes without external gluons lead to vanishing contributions, we must only
consider the gluon-photon induced top-pair production and then modify POWHEG in such a way
that the subtraction terms for the QED divergence in the $gq$ channel can also be
constructed.
We therefore compute here explicitly the expression for $\gr{B}_2^{\mu\nu}$, where the subscript $2$
designates the photon leg (see Fig.\ \ref{fig:06}). Applying the above procedure then leads to
\bea
  \gr{B}^{\mu\nu}_2 &=& {8\pi^2  \alpha_s\alpha Q_t^{2} \over m_t^2 z_1^2 y_1^2}\left(\begin{pmatrix}p_1^{\mu}&p_2^{\mu}&p_3^\mu\end{pmatrix}\gr{A}_1\begin{pmatrix}p_1^{\nu}\\p_2^{\nu}\\p_3^{\nu}\end{pmatrix} - \gr{A}_2 g^{\mu\nu}\right)\,,
\eea
where
\bea
\gr{A}_1 &=& 
  \begin{pmatrix}8 z_1^2& 2 \gr{P}_2 z_1& - 8 \gr{P}_1 z_1\\
	2 \gr{P}_2 z_1& 4 (\gr{P}_1 - z_1)^2 z_1&  6 \gr{P}_1 z_1^2 - 4 z_1^3 - 2\gr{P}_1^2 (2 + z_1)\\
	-8 \gr{P}_1 z_1&  6 \gr{P}_1 z_1^2 - 4 z_1^3 - 2\gr{P}_1^2 (2 + z_1)& 8 \gr{P}_1^2
  \end{pmatrix}\,,\\
  \gr{A}_2 &=&  m_t^2 \gr{P}_3 (\gr{P}_1 - z_1) z_1\,,\\
  \gr{P}_1 &=& y_1+z_1\,,\\
  \gr{P}_2 &=& 2(y_1+z_1) + y_1^2\,,\\
  \gr{P}_3 &=& y_1^2 + z_1^2\,,\\
  y_1 &=& \left( 1-\frac{t}{m_t^2} \right)\quad {\rm and}\\
  z_1 &=& \left( 1-\frac{u}{m_t^2} \right)\,.
\eea
As for the color-correlated squared Born matrix element, the closure relation of
Eq.\ (\ref{eq:3.10}) is implemented in POWHEG as a consistency check.


\subsection{Implementation of the virtual corrections}

For the implementation in POWHEG, the virtual corrections must be put into the form
\bea
 \gr{V} &=& \gr{N}\frac{\alpha_S}{2\pi}\left[ \frac{1}{\varepsilon^{2}}a\gr{B}
 + \frac{1}{\varepsilon}\sum_{i,j}c_{ij}\gr{B}_{ij} +\gr{V}_{\mathrm{fin.}} \right]
 \label{eq:3.19}
\eea
with the normalization constant
\bea
 \gr{N} &=& \frac{(4\pi)^{\varepsilon}}{\Gamma(1-\varepsilon)}\left( \frac{\mu_r^2}{Q^2}
 \right)^{\varepsilon}\,.
 \label{eq:3.20}
\eea
General expressions for the coefficients $a$ and $c_{ij}$ can be found, e.g., in App.\ B of
Ref.\ \cite{Frederix:2009yq} and in Refs.\ \cite{Jezo:2013,Lyonnet:2014wfa}.
$\mu_r$ is the renormalization scale, and $Q$ is an arbitrary
scale first introduced by Ellis and Sexton \cite{Ellis:1985er} and identified in POWHEG with
$\mu_r$. The finite part $\gr{V}_{\rm fin.}$ is then obtained form our calculation of the virtual
corrections in Sec.\ \ref{sec:2.2}.

\subsection{Real corrections and QED divergences}

Like the Born contributions, the real-emission squared amplitudes have been
implemented in POWHEG for each individual flavor structure contributing to the
real cross section. As already stated above, the diagram in Fig.\ \ref{fig:05}
(a) is finite and does not involve any singular regions. The diagrams in Fig.\
\ref{fig:04} and Fig.\ \ref{fig:05} (b) have the same underlying Born structure
as the LO process $q\bar{q}\to t\bar{t}$, followed or preceded by singular QCD
splittings of quarks into quarks (and gluons) or of gluons into quarks (and
antiquarks), so that their singular regions are automatically identified by POWHEG.

The diagrams in Fig.\ \ref{fig:05} (c) and (d) involve, however, the photon-induced
underlying Born diagrams in Fig.\ \ref{fig:06}, preceded by a singular QED
splitting of a quark into a photon (and a quark). The corresponding QED singularities
were so far not treated properly in POWHEG. Only the singular emission of final-state
photons had previously been implemented in Version 2 of the POWHEG BOX in the context
of the production of single $W$ bosons \cite{Barze:2012tt} and the neutral-current
Drell-Yan process \cite{Barze':2013yca}.

We therefore also implemented the photon-induced Born structures in Fig.\ \ref{fig:06},
replaced the POWHEG subtraction for the QCD splitting of initial quarks into gluons (and
quarks), which doesn't occur in our calculation, by a similar procedure for the QED
splitting of initial quarks into photons (and quarks), and enabled in addition the POWHEG
flag for real photon emission, which then allows for the automatic factorization of the
initial-state QED singularity and the use of photonic parton densities in the proton.
Note that this also restricts the possible choices of PDF parametrizations, as photon PDFs
are provided in very few global fits.

\subsection{Validation}

Our implementation of the electroweak top-pair production with new gauge-boson contributions
has been added to the list of POWHEG processes under the name PBZp. It allows for maximal
flexibility with respect to the choices of included interferences between SM photons and
$Z$ bosons as well as $Z'$ bosons, the vector and axial-vector couplings of the latter, and
the choices of renormalization and factorization scales (fixed or running with
$\sqrt{p_T^2+m_t^2}$ or $s$) in addition to the standard POWHEG options.

The SM Born, real and $1/\varepsilon$-expansion of the virtual matrix elements have been
checked against those provided by MadGraph5\_aMC@NLO \cite{Alwall:2014hca} and GoSam
\cite{Cullen:2011ac}, respectively. After including the $Z'$-boson contributions, we checked our
full implementation with respect to the cancellation of UV and IR divergences. We validated,
in addition to the renormalization procedure described in Sec.\
\ref{sec:2.2}, the completeness relations for the color- and spin-correlated Born amplitudes
and performed the automated POWHEG checks of
the kinematic limits of the real-emission amplitudes. In particular, we have checked explicitly
that the variable describing the collinear QED singularity shows a regular behavior after
the implementation of our new QED subtraction procedure. Restricting ourselves again to the
SM, our total hadronic cross section with the $q\bar{q}$ initial state only could be shown to
fully agree with the results in MadGraph5\_aMC@NLO, which does not allow for a proper treatment
of the QED divergence in the $gq$ initial state.

As already discussed in the introduction, the production of $Z'$ bosons
decaying to top pairs has been computed previously in NLO QCD by Gao et al.\ in a factorized approach for purely vector- and/or axial-vector-like couplings as those of the SSM \cite{Gao:2010bb}.
They neglected, however, all SM interferences and quark-gluon initiated diagrams with the $Z'$
boson in the $t$-channel. We can reproduce their $K$-factors of 1.2 to 1.4
(depending on the $Z'$ mass) up to 2\%, if we reduce our calculation to their theoretical
set-up and employ their input parameters. In the independent NLO QCD calculation by Caola et al.\
\cite{Caola:2012rs}, the authors include also the
additional quark-gluon initiated processes and show that they reduce the $K$-factor by about 5
\%. However, they still do not include the additional SM interferences, which they claim to be
small for large $Z'$-boson masses. As we have discussed in detail, this is not always true due to
the logarithmically enhanced QED contributions from initial photons. If we exclude SM interferences
and the (factorizable) QCD corrections to the top-quark decay, we can also reproduce their $K$-factors.

\section{Numerical results}
\label{sec:4}

In this section, we present numerical results for electroweak top-quark pair production including
$Z'$-boson contributions at LO and NLO from our new POWHEG code \cite{Alioli:2010xd}, which
we coupled to the parton shower and hadronization procedure in PYTHIA 8 \cite{Sjostrand:2007gs}.
Our results pertain to $pp$ collisions at the LHC with its current center-of-mass energy of
$\sqrt{S} = 13$ TeV. Only for total cross sections, we also study how much the reach in $Z'$
mass is extended in a future run at $\sqrt{S} = 14$ TeV. The top quark is assigned a
mass of $m_t = 172.5$ GeV as in the most recent ATLAS searches for $Z'$ bosons in this channel
\cite{Aad:2015fna} and is assumed to be properly reconstructed from its decay products.
At the top-pair production threshold, $\alpha(2m_t) = 1/126.89$. The
values of $\sin^2 \theta_W=0.23116$, $m_Z=91.1876$ GeV and $\Gamma_{Z} = 2.4952$ GeV were
taken from the Particle Data Group \cite{Agashe:2014kda}. The width of the $Z'$ boson depends
on its mass and its sequential Standard Model (SSM) or leptophobic topcolor
(TC) couplings. We vary the mass for total cross sections between 2 and 6 TeV and fix it
to 3 TeV for differential distributions.
 As stated in Sec.\ \ref{sec:1}, in the case of TC the $Z'$ width is set to 1.2\% of its mass,
 and the couplings are $f_1=1$ and $f_2=0$.
 We use the NNPDF23\_nlo\_as\_0118\_qed set of parton densities fitted with
$\alpha_{s}(m_Z)=0.118$, which includes the required photon PDF and allows
to estimate the PDF uncertainty \cite{Ball:2012cx,Ball:2013hta}.
The renormalization and factorization scales are varied by individual factors of two,
but excluding relative factors of four, around the central value $\mu_r=\mu_f=\sqrt{s}$.
In contrast to the two existing NLO calculations \cite{Gao:2010bb,Caola:2012rs}, which take only
the $Z'$-boson exchange and no SM interferences into account and where $m_{Z'}$ was chosen as the
central scale, our choice of $\sqrt{s}$ also applies to the SM channels and interpolates between
the different physical scales appearing in the process.

\subsection{Total cross sections}

\begin{figure}[!h]
 \centering
 \includegraphics[width=\textwidth]{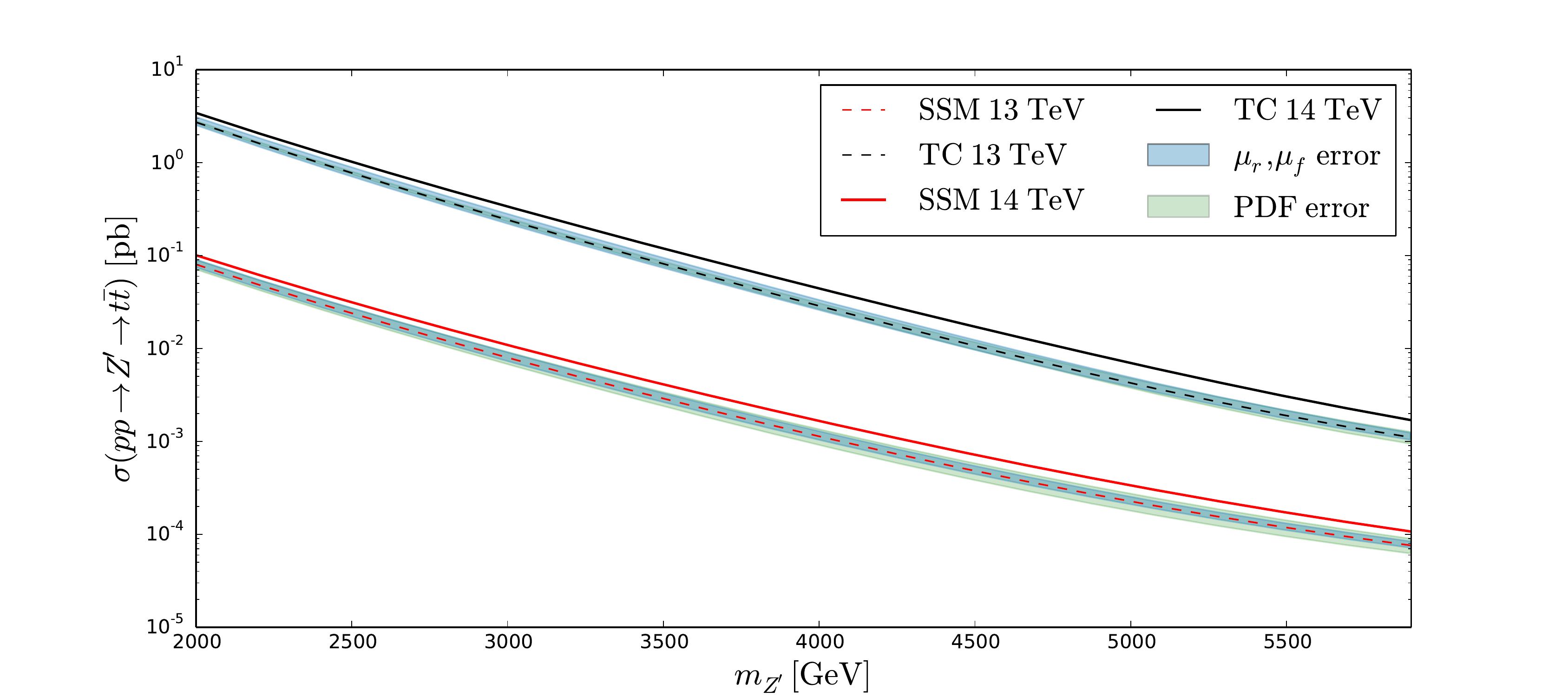}
 \caption{Total cross sections for $pp\to Z'\to t\bar{t}$ at the LHC with $\sqrt{S}=13$ TeV
 (dashed lines) and 14 TeV (full lines) as a function of the $Z'$ mass in NLO QCD for the
 sequential SM (SSM, red) and leptophobic topcolor model (TC, black). For $\sqrt{S}=13$ TeV,
 we also show the associated scale (blue) and PDF uncertainties (green) (color online).}
 \label{fig:07}
\end{figure}

To illustrate the total number of events to be expected from resonant-only $Z'$-boson
production at the LHC, we show in Fig.\ \ref{fig:07} the total NLO cross sections at a
center-of-mass energy of $\sqrt{S}=13$ TeV
in the SSM (dashed red curve) and TC (dashed black curve), together with the associated
renormalization and factorization scale uncertainties (blue bands) and PDF uncertainties
(green bands).
As one can see, in the case of the SSM (lower curves) the PDF uncertainty is larger than the
scale uncertainty in the entire range of $m_{Z'}$ masses from 2 to 6 TeV considered here.
Conversely, for the TC model (upper curves), it is the scale uncertainty which dominates for
$m_{Z'} \lesssim 5$ TeV, while the PDF uncertainty takes over only at larger values of
$m_{Z'}$, since the PDFs at large momentum fractions $x_{a,b}$ are less precisely known.
The uncertainties at NLO (note that the PS don't
affect the total cross sections) are about $\pm15$\% at low masses and increase to $\pm$35\%
in the SSM and $\pm20$\% in TC at higher masses. For an integrated luminosity of 100 fb$^{-1}$,
the number of expected
events falls from 10$^4$ for $m_{Z'}=2$ TeV to 10 for $m_{Z'}=6$ TeV
in the SSM and is about an order of magnitude larger in TC.
When the LHC energy is increased to 14 TeV, the corresponding total cross sections (full
curves) at high $Z'$-boson mass are larger by about 50\%, and the mass reach is extended by about
500 GeV, less of course than the increase in the hadronic energy $\sqrt{S}$, of which
only a fraction is transferred to the initial partons and the hard scattering.

\begin{figure}[t]
 \centering
 \includegraphics[scale=0.35]{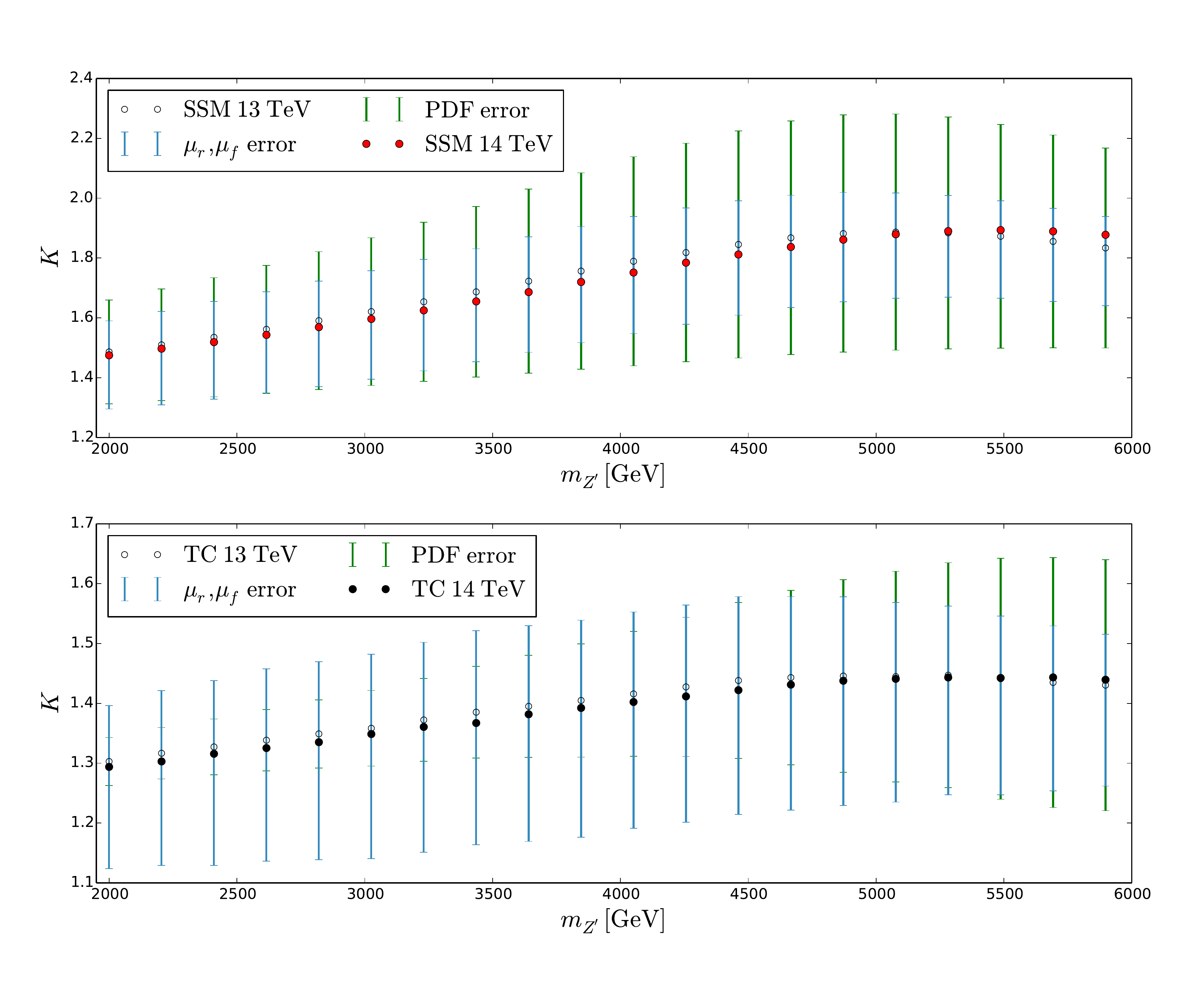}
 \caption{$K$-factors (i.e.\ ratios of NLO/LO cross sections) at the LHC with $\sqrt{S}=13$ TeV
 (open circles) and 14 TeV (full circles) as functions of the $Z'$ mass for the SSM (top) and TC
 (bottom). For $\sqrt{S}=13$ TeV, we also show the associated scale (blue) and PDF uncertainties
 (green) (color online).}
 \label{fig:08}
\end{figure}

Even for resonant-only $Z'$-boson production, the $K$-factor is not completely mass-independent,
as can be seen in Fig.\ \ref{fig:08}. In TC (lower plot), it increases only modestly from 1.3 to
1.45 in the
mass range considered here, while in the SSM (upper plot) it increases much more from about 1.45
to 1.85. In contrast, it depends very little on the LHC center-of-mass energy of 13 TeV (open
circles) or 14 TeV (full circles). In this figure, the scale and PDF uncertainties can also
be read off more precisely than in the previous figure. \\

In Tab.\ \ref{tab:01} we list the total cross sections in LO for top-pair production at
\begin{table}
\caption{\label{tab:01}Total cross sections in LO for top-pair production at
 ${\cal O}(\alpha_s^2)$, ${\cal O}(\alpha_s\alpha)$ and ${\cal O}(\alpha^2)$
 in the SM, SSM and TC, together with the corresponding NLO corrections.
 The $Z'$-boson mass is set to 3 TeV.}\vspace*{3mm}
\begin{tabular}{lllrr}
Order&Processes& Model & $\sigma$ [pb] & $\sigma$ [pb] $(m_{t\bar{t}}>{3\over4}m_{Z'})$\\
\hline
\hline
LO& $q\bar{q}/gg\to t\bar{t}$            && 473.93(7)         &   0.15202(2)\\
NLO&$q\bar{q}/gg+qg\to t\bar{t}+q$               &&1261.0(2)          &   0.45255(7)\\
\hline
LO&$\gamma g+g\gamma     \to t\bar{t}$   &&   4.8701(8)       &   0.0049727(6)\\
LO &$\gamma g+g\gamma \to t\bar{t}$ \hfill (NLO $\alpha_s$ and PDFs) &&   5.1891(8)       &   0.004661(6)\\
\hline
LO&$q\bar{q}\to \gamma/Z\to t\bar{t}$    &SM&   0.36620(7)      &   0.00017135(3)\\
NLO&$q\bar{q}\to \gamma/Z\to t\bar{t}$   &SM&   0.5794(1)       &   0.00017174(5)\\
NLO&$q\bar{q}+qg\to \gamma/Z+q\to t\bar{t}+q$  &SM&   4.176(2)        &   0.001250(6)\\
\hline
LO&$q\bar{q}\to Z' \to t\bar{t}$  &SSM   &   0.0050385(8)    &   0.0044848(7)\\
LO&$q\bar{q}\to \gamma/Z/Z'\to t\bar{t}$ &SSM &   0.35892(7)      &   0.0043464(7)\\
NLO&$q\bar{q}\to \gamma/Z/Z'\to t\bar{t}$ & SSM    &   0.5676(1)       &   0.005155(3)\\
NLO&$q\bar{q}+qg\to \gamma/Z/Z'+q\to t\bar{t}+q$ &SSM    &   4.172(2)        &   0.007456(9)\\
\hline
LO&$q\bar{q}\to Z' \to t\bar{t}$ &TC    &   0.012175(2)     &   0.011647(2)\\
LO&$q\bar{q}\to \gamma/Z/Z'\to t\bar{t}$ &TC  &   0.38647(7)      &   0.011984(2)\\
NLO&$q\bar{q}\to \gamma/Z/Z'\to t\bar{t}$ &TC  &   0.6081(2)       &   0.01468(1)\\
NLO&$q\bar{q}+qg\to \gamma/Z/Z'+q\to t\bar{t}+q$ &TC  &   4.202(2)        &   0.01002(1)\\
\end{tabular}
\end{table}
${\cal O}(\alpha_s^2)$, ${\cal O}(\alpha_s\alpha)$ and ${\cal O}(\alpha^2)$
in the SM, SSM and TC, i.e.\ including the SM backgrounds, together with the
corresponding NLO corrections. The $Z'$-boson mass is set here to 3 TeV,
and for our LO predictions we use  the NNPDF23\_lo\_as\_0119\_qed PDF set,
since a set with $\alpha_s(m_Z)=0.118$ is not available at this order.
Comparing first the LO results only, we observe that the pure QCD processes of
${\cal O}(\alpha_s^2)$ have a total cross section of about 474 pb, i.e.\
two orders of magnitude larger than the photon-gluon induced processes of
${\cal O}(\alpha_s\alpha)$ with 4.87 pb as naively expected from the ratio of
strong and electromagnetic coupling constants in the hard scattering and in
the PDFs. The suppression of the pure electroweak with respect to QCD processes is
more than three orders of magnitude, as expected from the ratio of
coupling constants in the hard scattering and when taking into account
that the QCD processes have both quark- and gluon-initiated contributions.
The $Z'$-mediated processes in the SSM and TC have only cross sections of
5 and 12 fb, respectively compared to 366 fb from the SM channels alone,
which therefore clearly dominate the total electroweak cross sections.
The interference effects are destructive in the SSM ($-4$\%), but constructive
in TC ($+2$\%).

When a cut on the invariant mass of the top-quark pair of 3/4 of the $Z'$
mass (i.e.\ at 2.25 TeV) is introduced, the SM backgrounds are reduced by more than
three orders of magnitude, while the signal cross sections drop only by
about 10\%. The interference effects then become more important in
the SSM ($-7$\%), but not in TC ($+2$\%) with its very narrow $Z'$ width of 1.2\%
of its mass. While an invariant-mass cut strongly enhances the signal-to-background
ratio, the LHC experiments still have to cope with signals that reach only
3 to 8 \% of the QCD background, which makes additional cuts on kinetic variables
necessary.

The NLO corrections for the QCD processes are well-known and can be computed
with the published version of POWHEG (HVQ) \cite{Frixione:2007nw}.
At the LHC with its high gluon luminosity,
the $qg$ channels opening up at NLO are known to introduce large $K$-factors,
here of about a factor of three. The NLO corrections for the purely electroweak processes
are new even in the SM, where we have introduced a proper subtraction procedure
for the photon-induced processes. The $K$-factors for the $q\bar{q}$ channel
are moderate in the SM (+56\%), SSM (+58\%) and TC (+56\%), where the last
two numbers are dominated by SM contributions and therefore very similar.
Only after the invariant-mass cut the differences in the models become more
apparent in the $K$-factors for the SM ($\pm0$\%), SSM ($+19$\%) and TC ($+23$\%).
However, similarly to the QCD case the $qg$ channel,
and also the $\gamma g$ channel opening up for the first time at this order,
introduce contributions much larger than the underlying Drell-Yan type Born
process. Note that the LO $\gamma g$ cross section computed with NLO
$\alpha_s$ and PDFs must still be added to the full NLO $q\bar{q}+gg$ cross
sections. An invariant-mass cut is then very instrumental to bring down
the $K$-factors and enhance perturbative stability, as one can see from the
LO $\gamma g$ and in particular the NLO results in the SSM and TC.

\subsection{Differential distributions}

We now turn to differential cross sections for the electroweak production of top-quark
pairs that includes the contribution of a SSM or TC $Z'$ boson with a fixed mass of 3 TeV.

\begin{figure}[!h]
 \centering
 \includegraphics[scale=0.65]{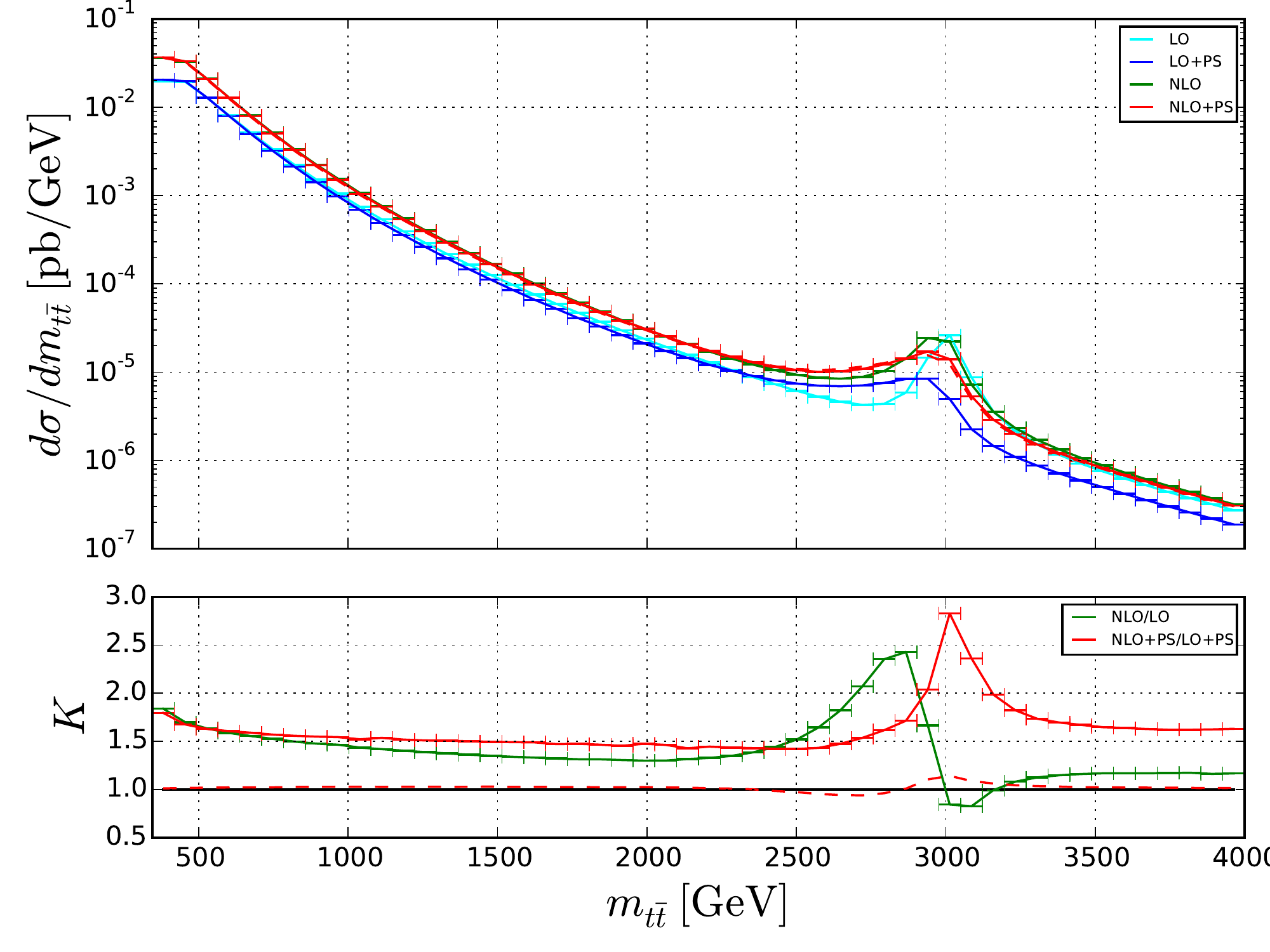}
 \includegraphics[scale=0.65]{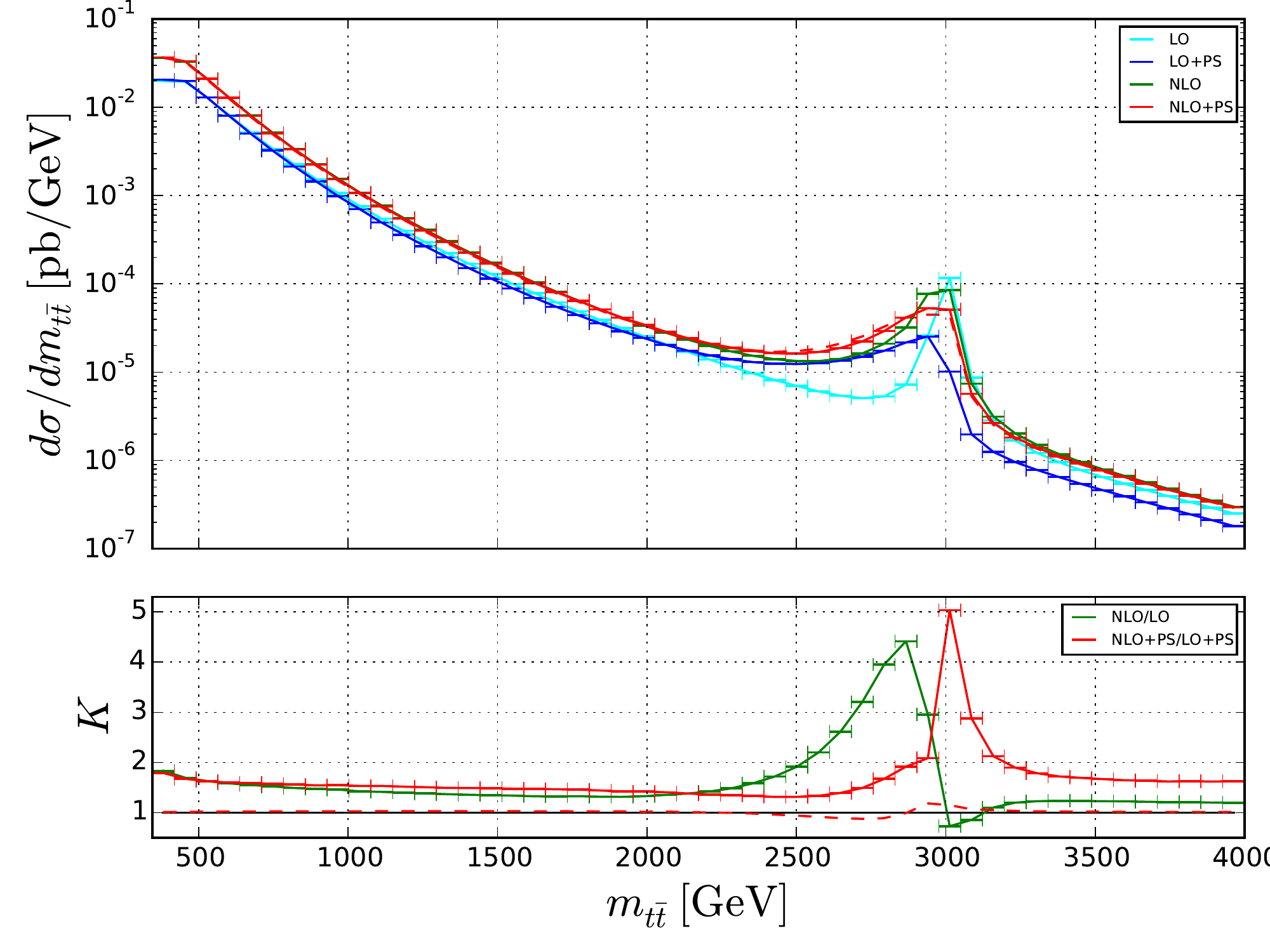}
 \caption{Invariant-mass distributions of top-quark pairs produced through $\gamma$, $Z$ and
 $Z'$ bosons and their interferences at the LHC with $\sqrt{S}=13$ TeV at LO (light blue),
 LO+PS (dark blue), NLO (green) and NLO+PS (red) accuracy together with the corresponding
 $K$-factors in the SSM (top) and TC (bottom). The dashed red curves have been obtained
 with HERWIG 6 \cite{Corcella:2000bw} instead of PYTHIA 8 \cite{Sjostrand:2007gs} (color online).}
 \label{fig:09}
\end{figure}

The invariant-mass distributions of top-quark pairs in Fig.\ \ref{fig:09} exhibit
steeply falling spectra from the SM background from 10$^{-2}$ to 10$^{-7}$ pb/GeV together
with clearly visible resonance peaks of SSM (top) and TC (bottom) $Z'$ bosons at 3 TeV,
whose heights and widths differ of course due to the different couplings to SM particles
in these two models. In particular, the TC resonance cross section is about an order of
magnitude larger than the one in the SSM in accordance with the total cross section results in
the previous subsection (see Fig.\ \ref{fig:07}).
What becomes also clear from the lower panels in Fig.\ \ref{fig:09} (top and bottom) is that
the $K$-factors are highly dependent on the invariant-mass region and can reach large
factors around the resonance region. This is particularly true for TC (bottom), but also
for the SSM, and related to the fact that the position of the
resonance peak is shifted towards lower invariant masses from LO to NLO due to additional
radiation at this order. As one can see, this effect is already present if parton showers
are added to the LO calculation, so that the NLO+PS to LO+PS comparison mostly results in
an increased $K$-factor at and above the resonance.

\begin{figure}[!h]
 \centering
 \includegraphics[scale=0.65]{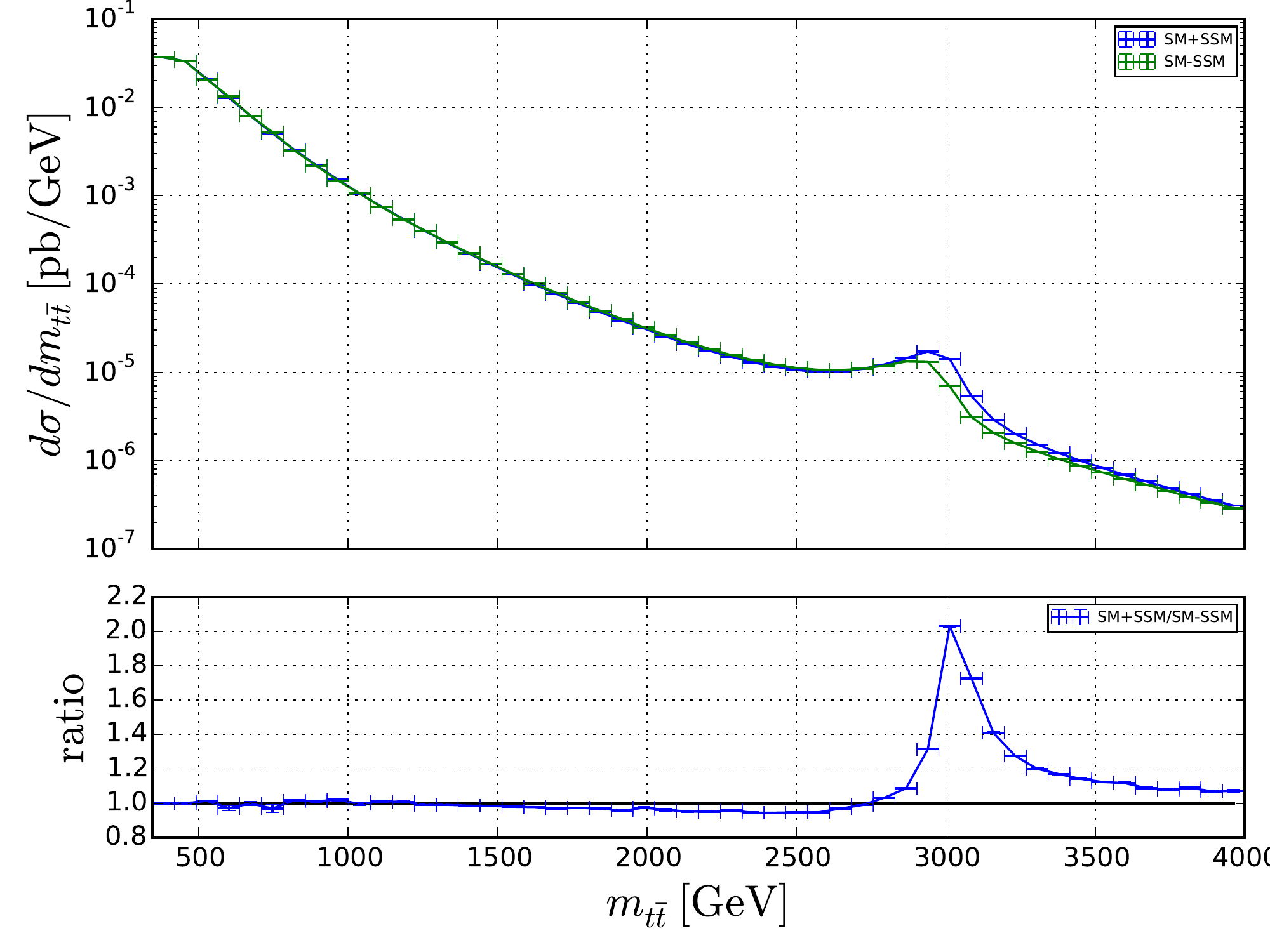}
 \includegraphics[scale=0.65]{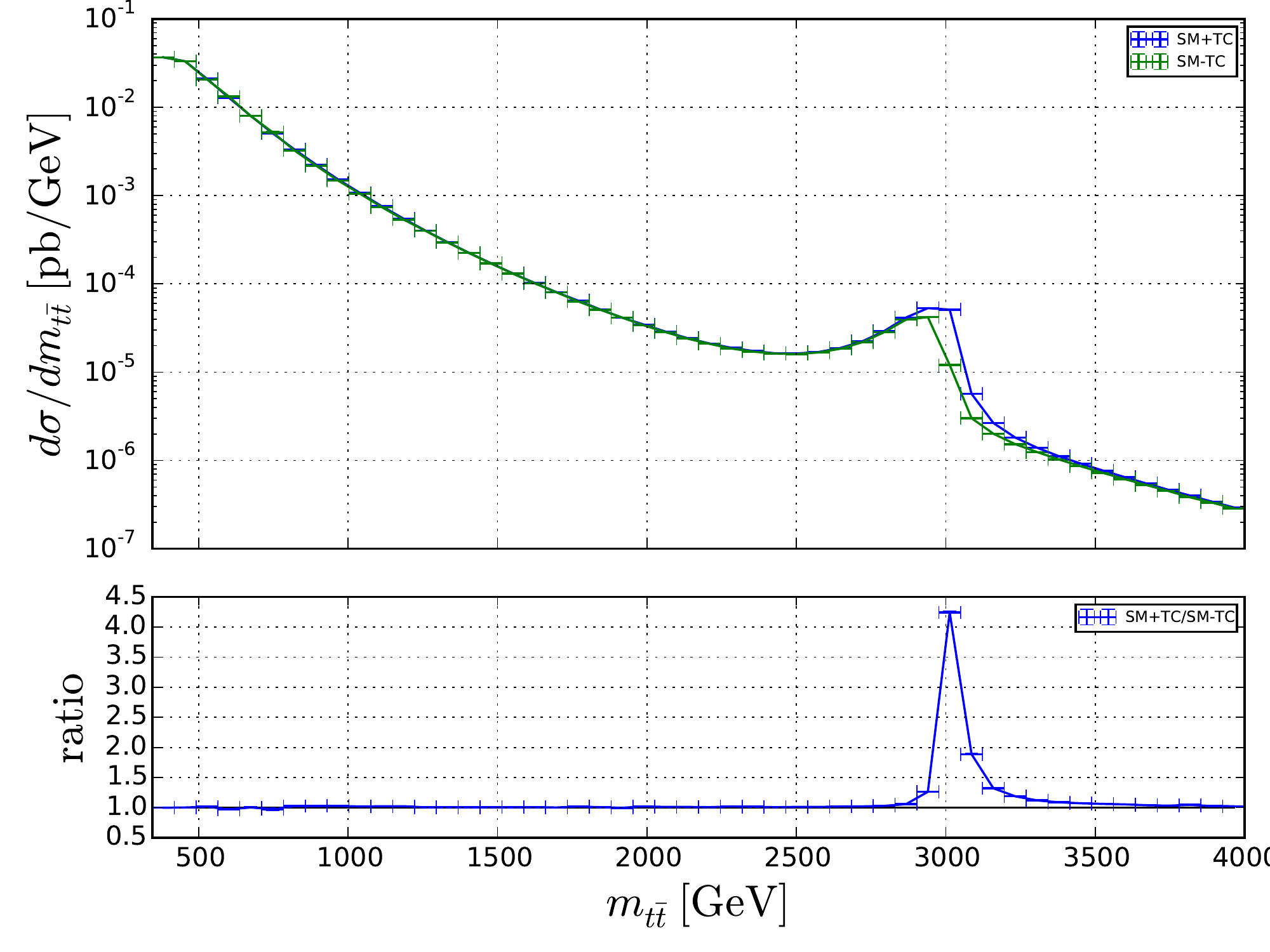}
 \caption{Invariant-mass distributions of top-quark pairs produced through $\gamma$, $Z$ and
 $Z'$ bosons with (green) and without interferences (blue) at the LHC with
 $\sqrt{S}=13$ TeV at NLO+PS accuracy together with the corresponding ratios in the SSM (top)
 and TC (bottom) (color online).}
 \label{fig:10}
\end{figure}

The effect of interferences between SM and new physics contributions is shown in
Fig.\ \ref{fig:10}, where the sum of the squared individual contributions (blue)
is compared with the square of the sum of all contributions (green) in the SSM
(top) and TC (bottom). As one can see, the interference effects shift the resonance
peaks to smaller masses, and their sizes are reduced. When the ratios
of the two predictions are taken (lower panels), it becomes clear that predictions
without interferences overestimate the true signal by a factor of two or more.

\begin{figure}[!h]
 \centering
 \includegraphics[scale=0.65]{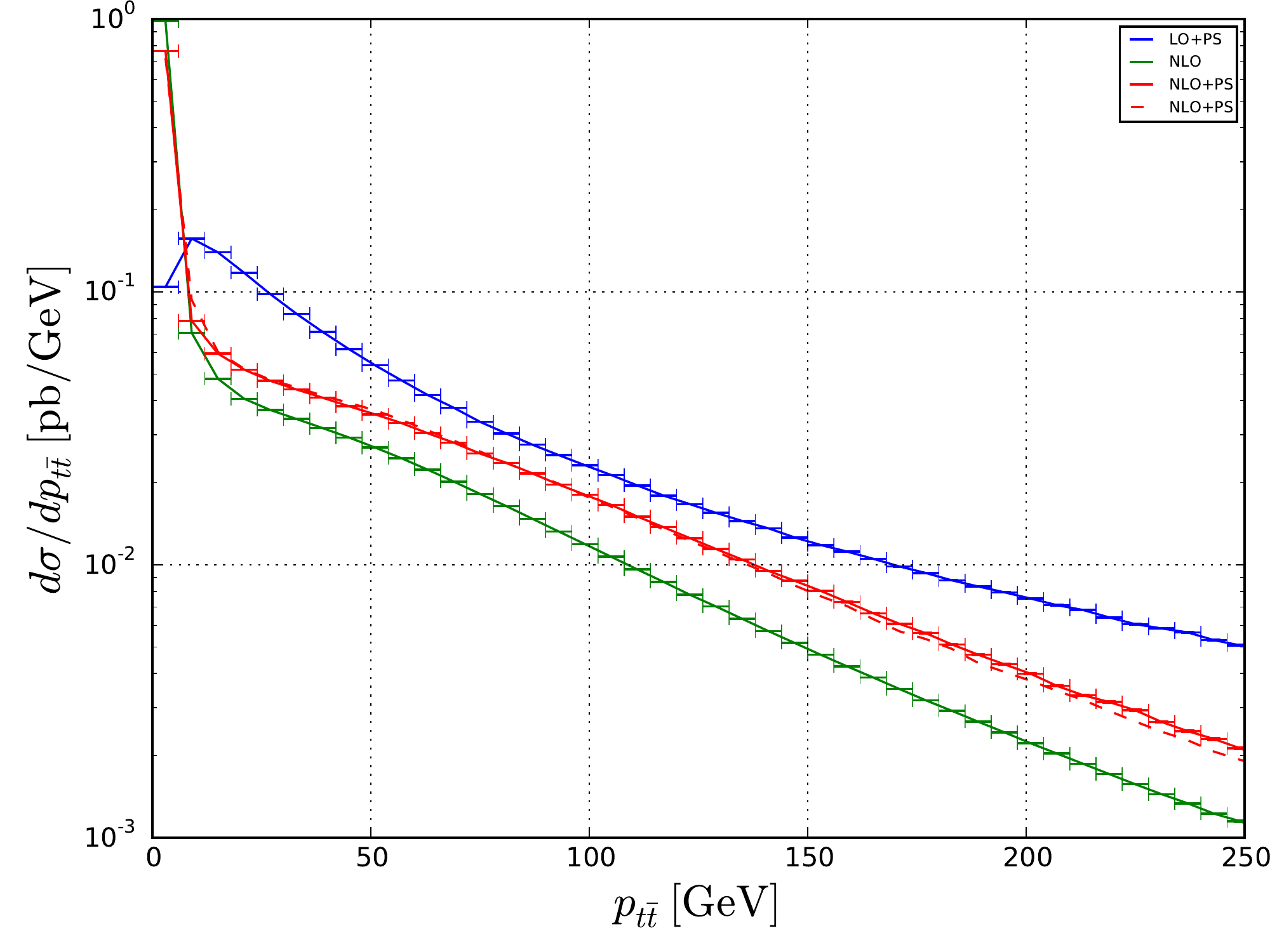}
 \includegraphics[scale=0.65]{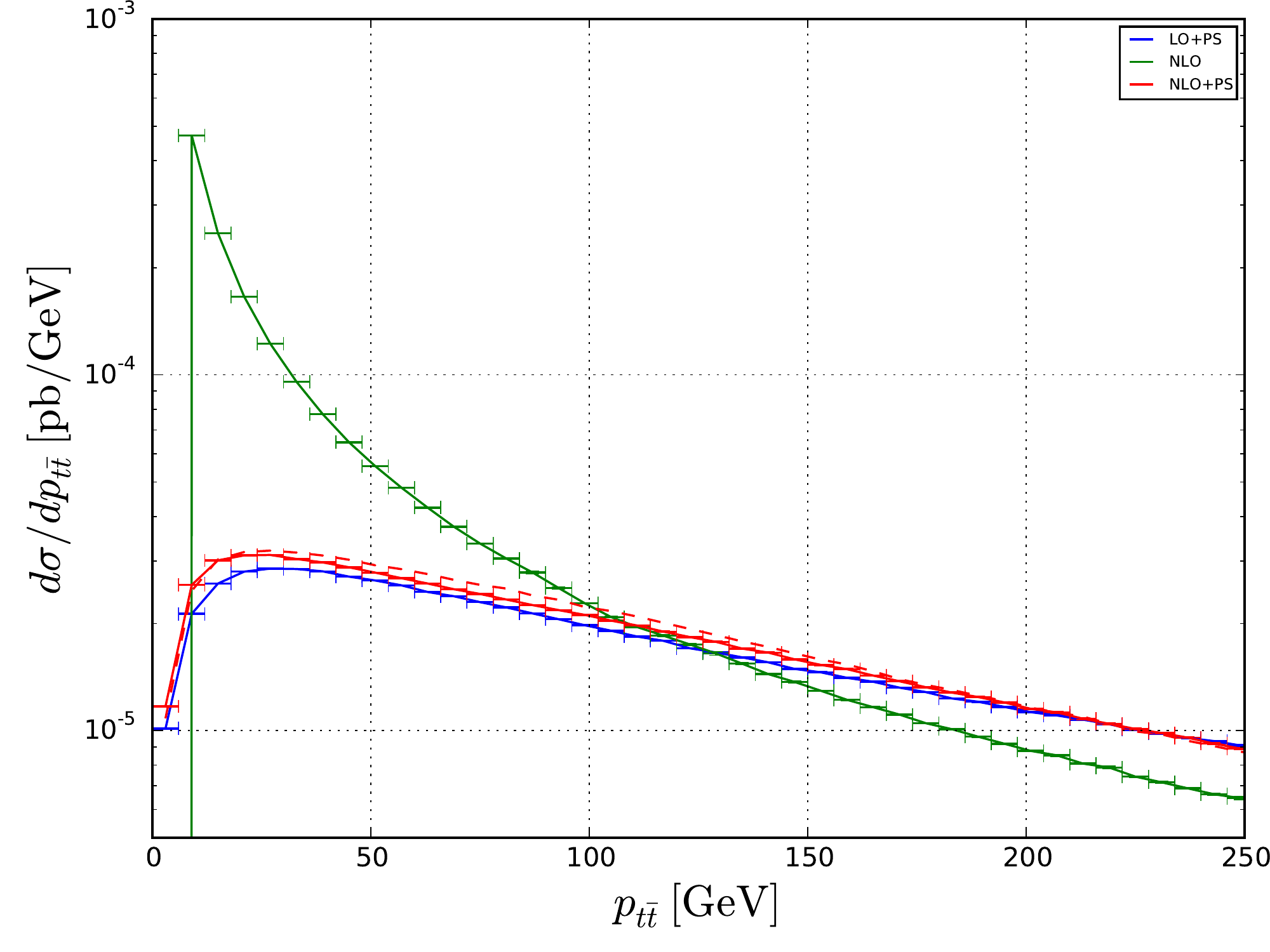}
 \caption{Transverse-momentum distributions of top-quark pairs produced through $\gamma$, $Z$ and
 $Z'$ bosons and their interferences (top) and through $Z'$ bosons alone (bottom) at the LHC with
 $\sqrt{S}=13$ TeV at LO+PS (dark blue), NLO (green) and NLO+PS (red) accuracy in the SSM.
 The TC distributions look very similar.  The dashed red curves have been obtained
 with HERWIG 6 \cite{Corcella:2000bw} instead of PYTHIA 8 \cite{Sjostrand:2007gs} (color online).}
 \label{fig:11}
\end{figure}

\begin{figure}[!h]
 \centering
 \includegraphics[scale=0.65]{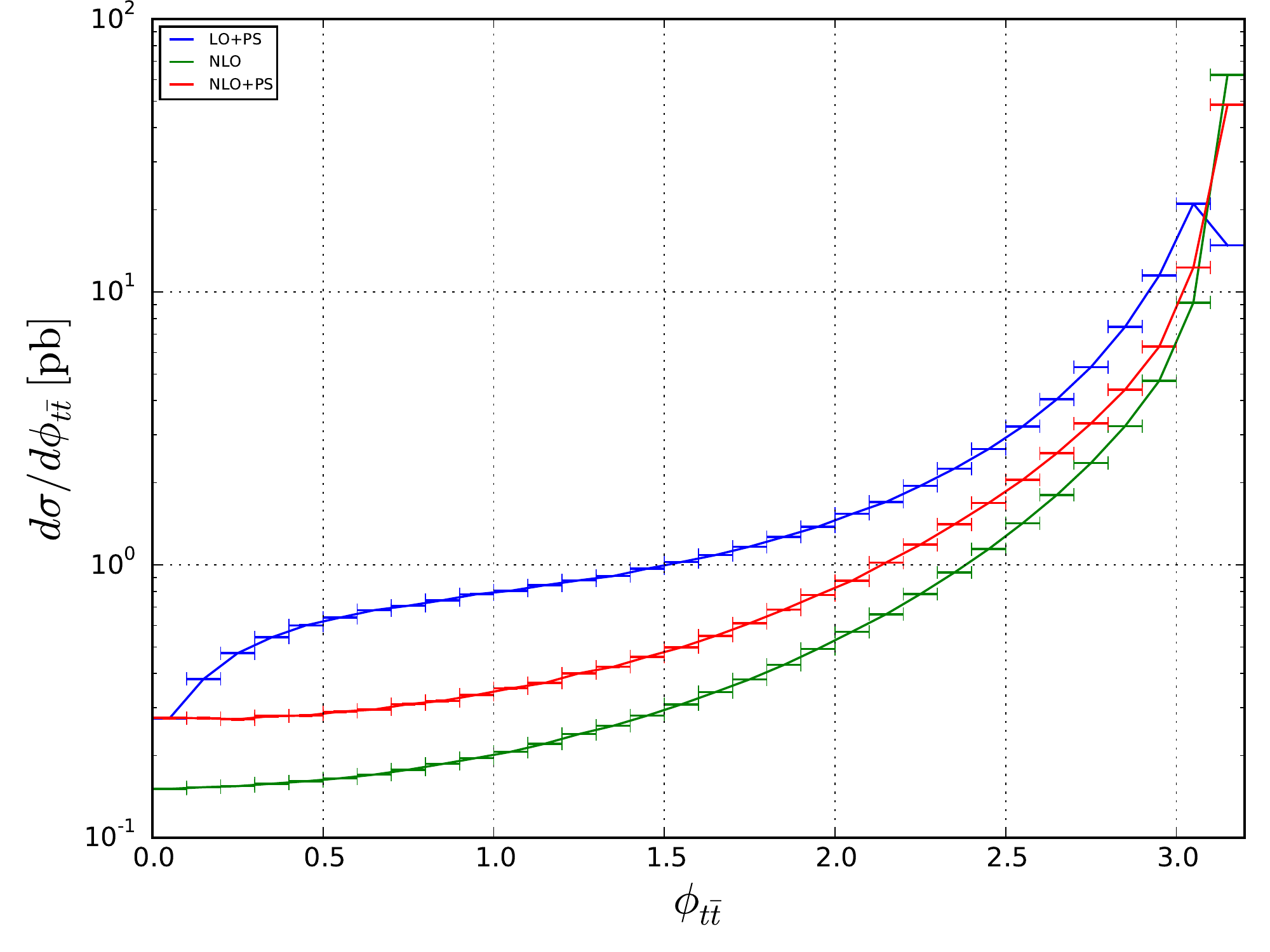}
 \includegraphics[scale=0.65]{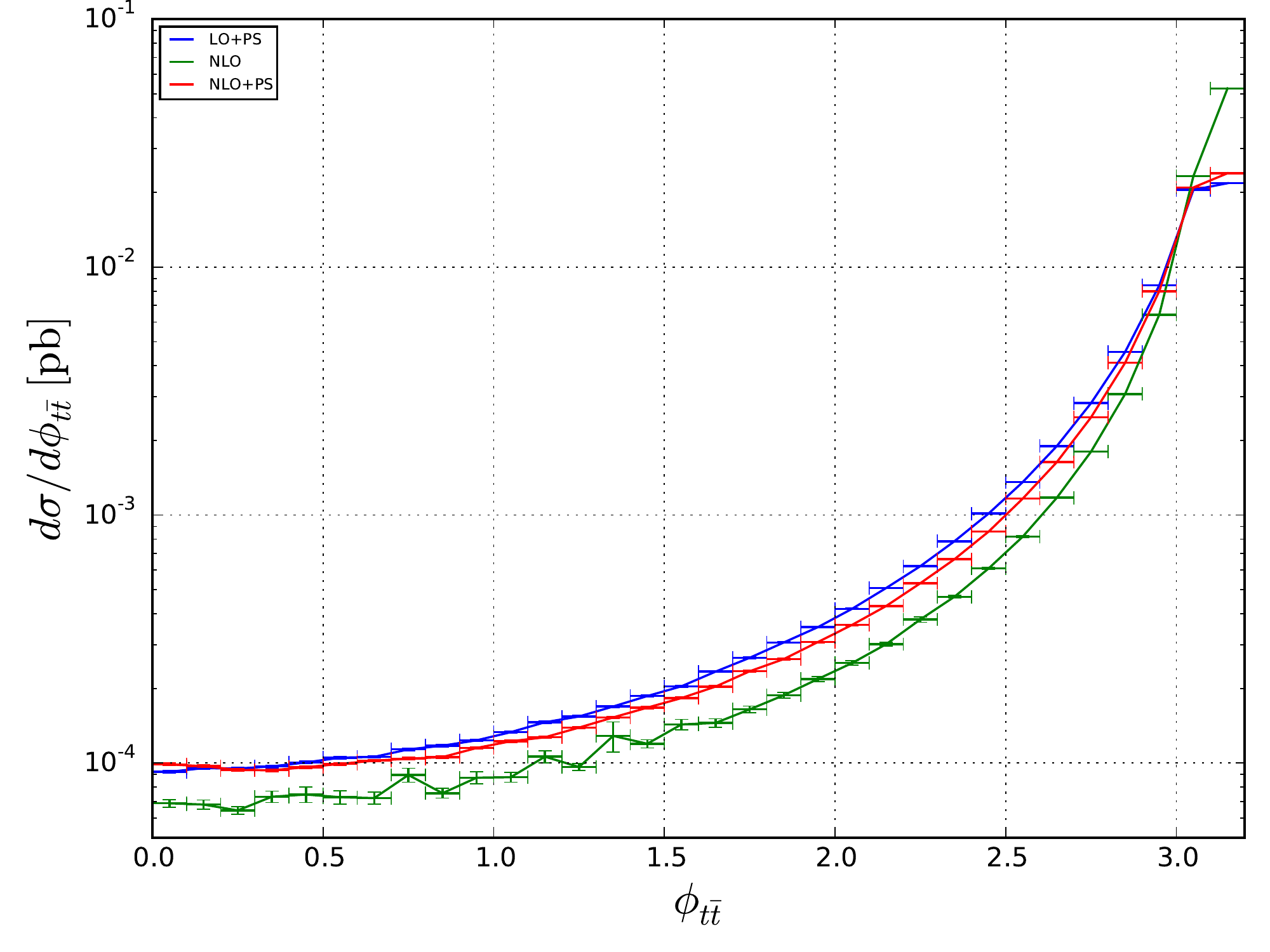}
 \caption{Distributions in the azimuthal opening angle of top-quark pairs produced through
 $\gamma$, $Z$ and $Z'$ bosons and their interferences (top) and through $Z'$ bosons alone
 (bottom) at the LHC with $\sqrt{S}=13$ TeV at LO+PS (dark blue), NLO (green) and NLO+PS
 (red) accuracy in the SSM. The TC distributions look very similar (color online).}
 \label{fig:12}
\end{figure}

The two variables that are particularly sensitive to soft-parton radiation and the
associated resummation in NLO+PS Monte Carlo programs are the net transverse momentum
of the observed particle (here top-quark) pair ($p_{t\bar{t}}$) and the azimuthal opening
angle between them ($\phi_{t\bar{t}}$), which are 0 and $\pi$, respectively, at LO.
At NLO they are
balanced by just one additional parton and thus diverge and exhibit physical
behavior and turnover only at NLO+PS, i.e.\ after resummation of the left-over kinematical
singularities. These well-known facts can also be observed in Figs.\ \ref{fig:11} and
\ref{fig:12}, where for obvious reasons the LO $\delta$-distributions at 0 and $\pi$
are not shown. As expected, the NLO (green) predictions diverge close to these end points,
while the NLO+PS (red) predictions approach finite asymptotic values. Again, a similar
behavior is already observed at LO+PS accuracy, although with different normalization and
shape. Interestingly, the resummation works much better for purely $Z'$-mediated processes
(lower panels) than if SM and interference contributions are included (upper panels).
This effect can be traced back to the fact that in the SM-dominated full cross section the
top-pair production threshold at $2m_t=345$ GeV is almost one order of magnitude smaller than
the mass $m_{Z'}=3$ TeV governing the exclusive $Z'$-boson channel.

%

 In our discussion of total cross sections in Sec.\ 4.1, we had
 included analyses of scale and PDF uncertainties at NLO, but
 not of the uncertainty coming from different PS implementations,
 as the PS does not influence total cross sections, but only
 differential distributions. To estimate this uncertainty, we
 therefore show in Figs.\ \ref{fig:09} and \ref{fig:11} also results
 obtained with the HERWIG 6 PS (dashed red) \cite{Corcella:2000bw} in
 addition to those obtained with our standard PYTHIA 8 PS (full red)
 \cite{Sjostrand:2007gs}. The dashed red curves in the lower
 panels of Fig.\ \ref{fig:09} represent the ratios of the HERWIG 6
 over the PYTHIA 8 PS results. As one can see there, the invariant-mass
 distributions in the SSM and TC are enhanced by the HERWIG 6 PS at the
 resonance at 3 TeV by about
 10\%, while the region just below it is depleted by a smaller
 amount, but over a larger mass region. The PS differences are therefore
 smaller (by factors of three to six, except for the PDF error in TC)
 than those of the scale and PDF uncertainties in Fig.\ \ref{fig:08}.
 The SSM transverse-momentum distribution in Fig.\ \ref{fig:11} falls
 off a bit faster with the HERWIG 6 PS than with the PYTHIA 8 PS at large
 transverse momenta, while in TC it is slightly enhanced at low values,
 but no significant differences appear between the angularly ordered
 HERWIG 6 PS and the dipole PS in PYTHIA 8.

 The importance of next-to-leading-logarithmic (NLL) contributions
 that go beyond the leading-logarithmic (LL) PS accuracy can be
 estimated by a comparison with analytic NLL resummation calculations.
 These have not been performed for top-quark, but only for lepton final
 states \cite{Fuks:2007gk}. In Fig.\ 5 of this paper, it has been
 found that the invariant-mass distribution shows no significant
 difference, while the LL transverse-momentum distribution computed with
 the HERWIG 6 PS is somewhat smaller than the one obtained with NLL resummation,
 but that it stays within the residual scale uncertainty of the latter.

\begin{figure}[!h]
 \centering
 \includegraphics[scale=0.65]{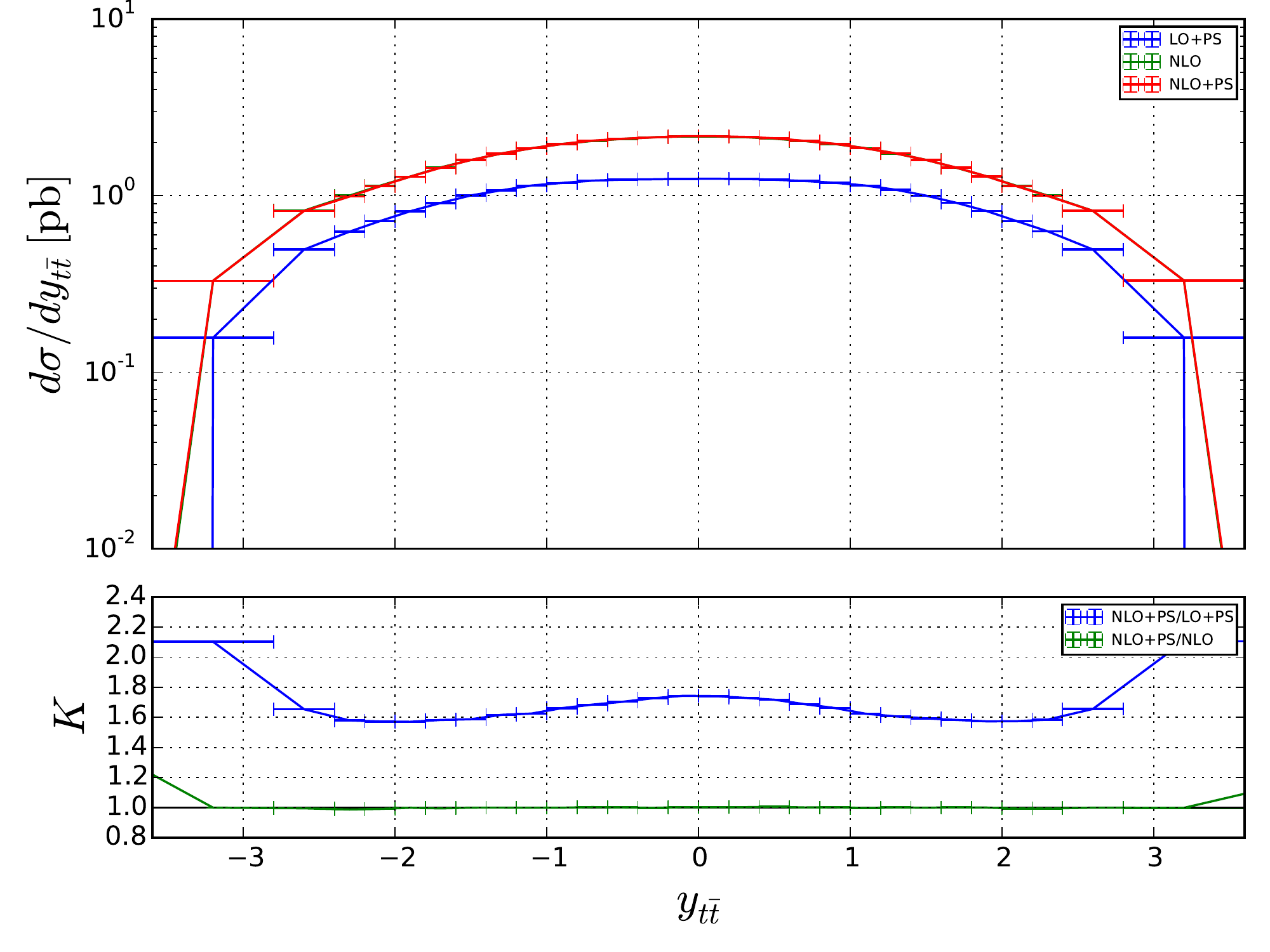}
 \includegraphics[scale=0.65]{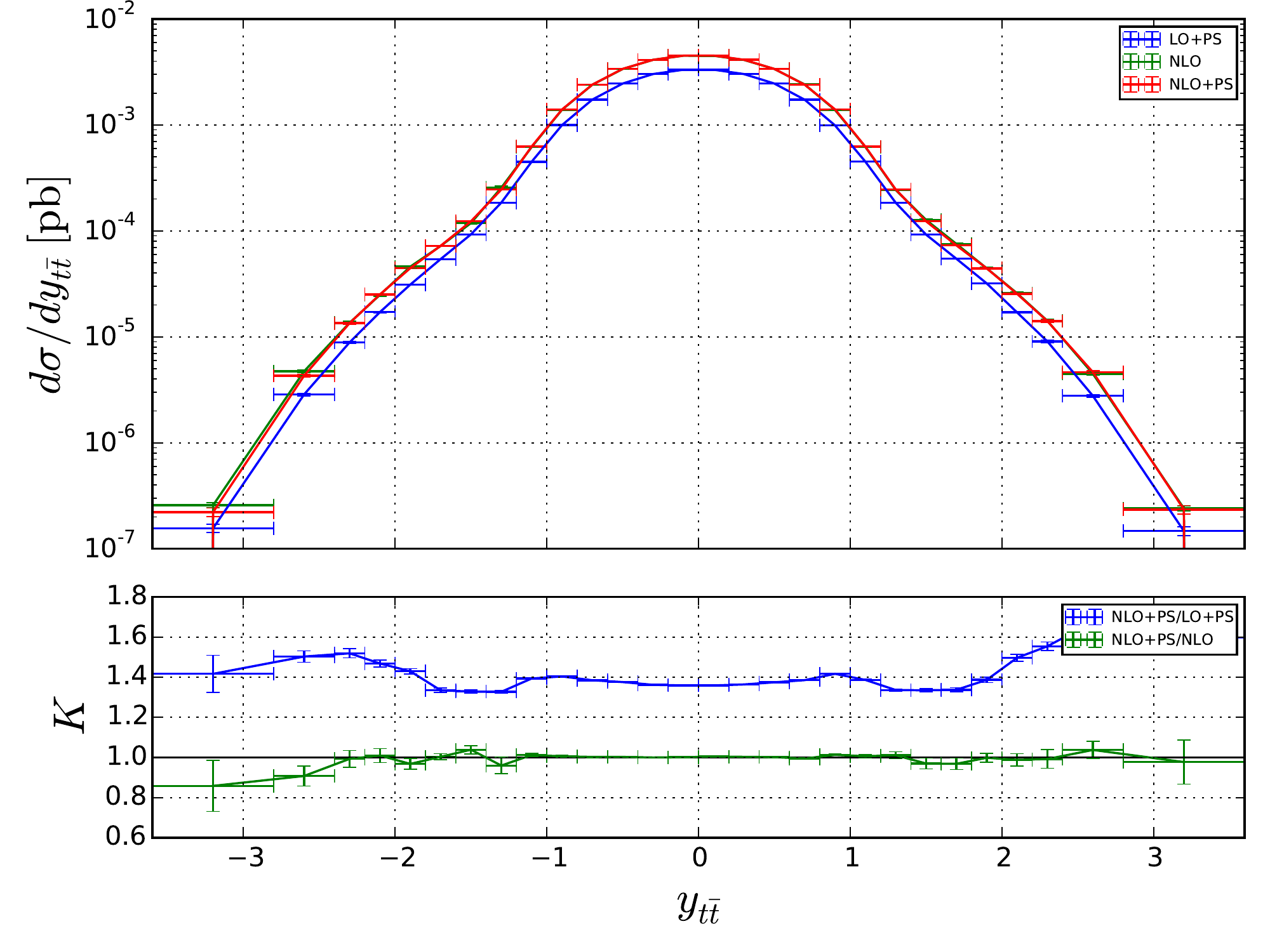}
 \caption{Rapidity distributions of top-quark pairs produced through
 $\gamma$, $Z$ and $Z'$ bosons and their interferences (top) and through $Z'$ bosons alone
 (bottom) at the LHC with $\sqrt{S}=13$ TeV at LO+PS (dark blue), NLO (green) and NLO+PS
 (red) accuracy together with the corresponding $K$-factors in the SSM. The NLO and NLO+PS
 curves nearly coincide here (color online).}
 \label{fig:13}
\end{figure}
\begin{figure}[!h]
 \centering
 \includegraphics[scale=0.65]{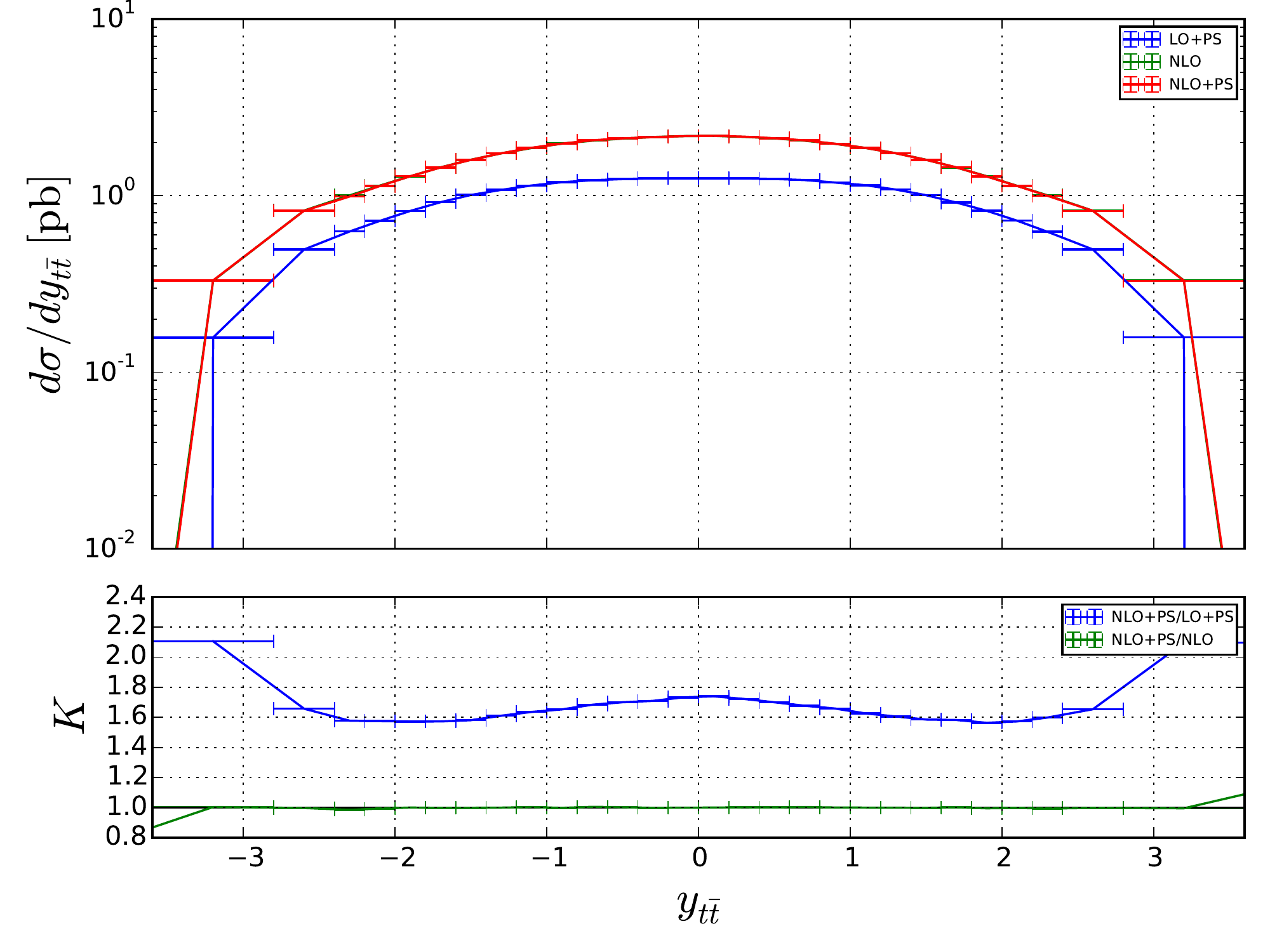}
 \includegraphics[scale=0.65]{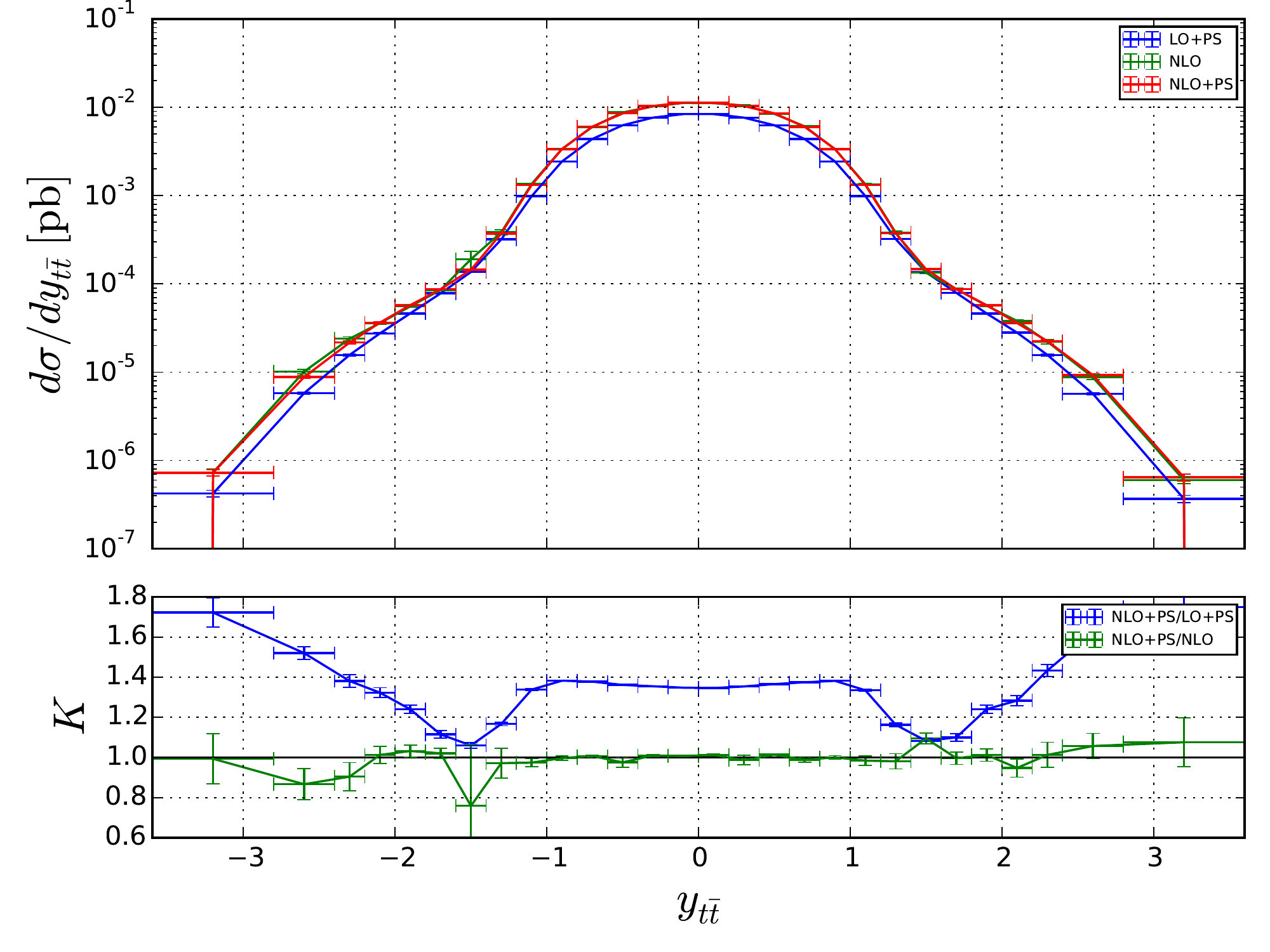}
 \caption{Same as Fig.\ \ref{fig:13}, but for TC (color online).}
 \label{fig:14}
\end{figure}

Rapidity distributions of the top-quark pair are shown in Figs.\ \ref{fig:13} and \ref{fig:14}.
If SM contributions are taken into account (top), they are much flatter than if only the heavy
resonance contributes (bottom), i.e.\ the top-quark pairs are then produced much more centrally.
The effect is similar, but somewhat less pronounced in TC (Fig.\ \ref{fig:14}) than in the SSM
(Fig.\ \ref{fig:13}) due to the broader resonance in this model. Even for rapidity distributions
NLO effects are not simply parametrizable by a global $K$-factor, as it varies from 1.6 to
2.1, when SM contributions are taken into
account (blue curves in the upper $K$-factor panels) and drops from 1.6 to 1.4 or even below,
if they are not taken into account (blue curves in the lower $K$-factor panels). As expected, the
parton showers (green curves in the $K$-factor panels) have little effect on the central parts of
the rapidity distributions, and they only slightly influence the forward/backward regions through
additional parton radiation from the initial state.


\begin{figure}[!h]
 \centering
 \includegraphics[scale=0.65]{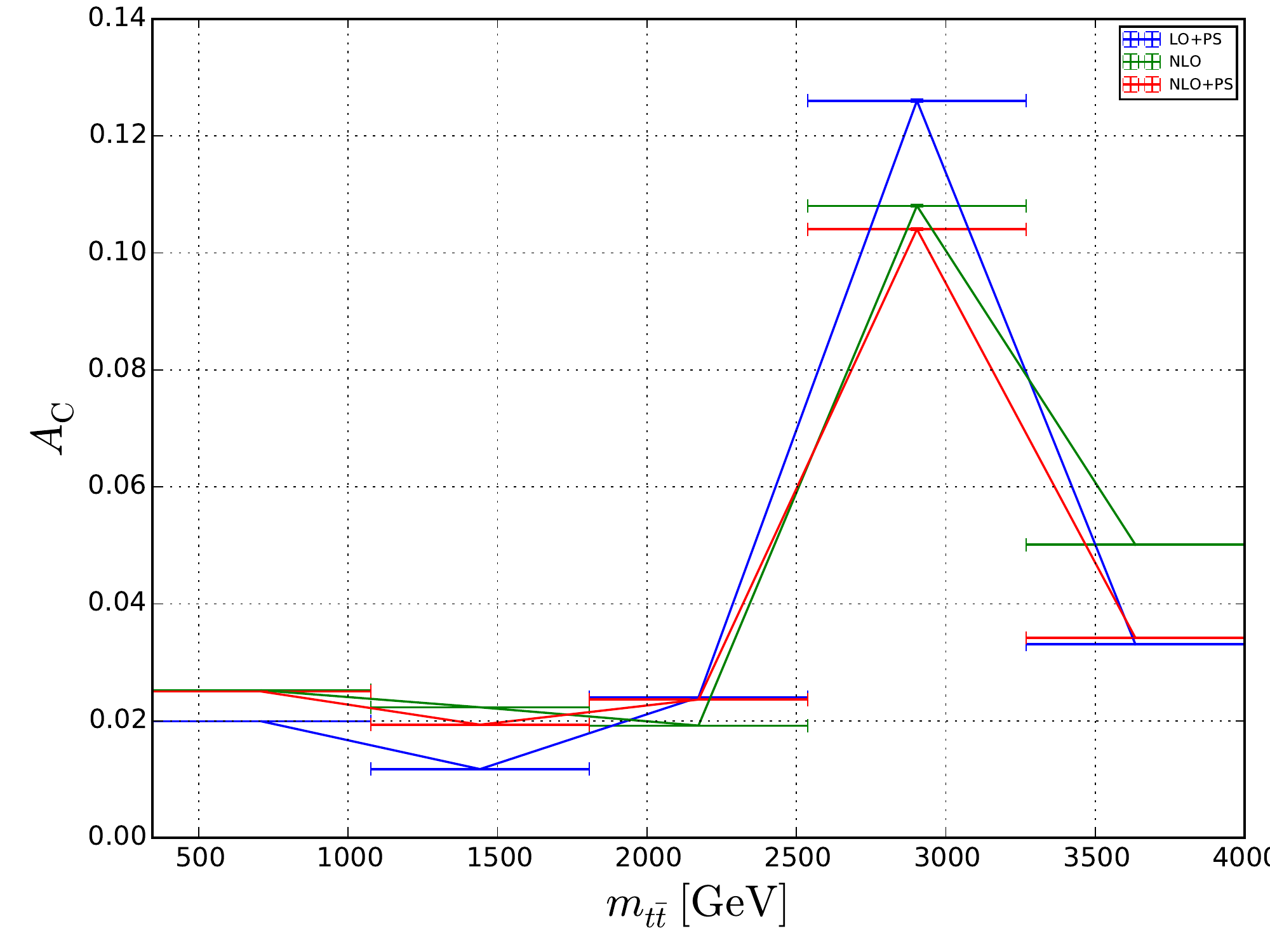}
 \includegraphics[scale=0.65]{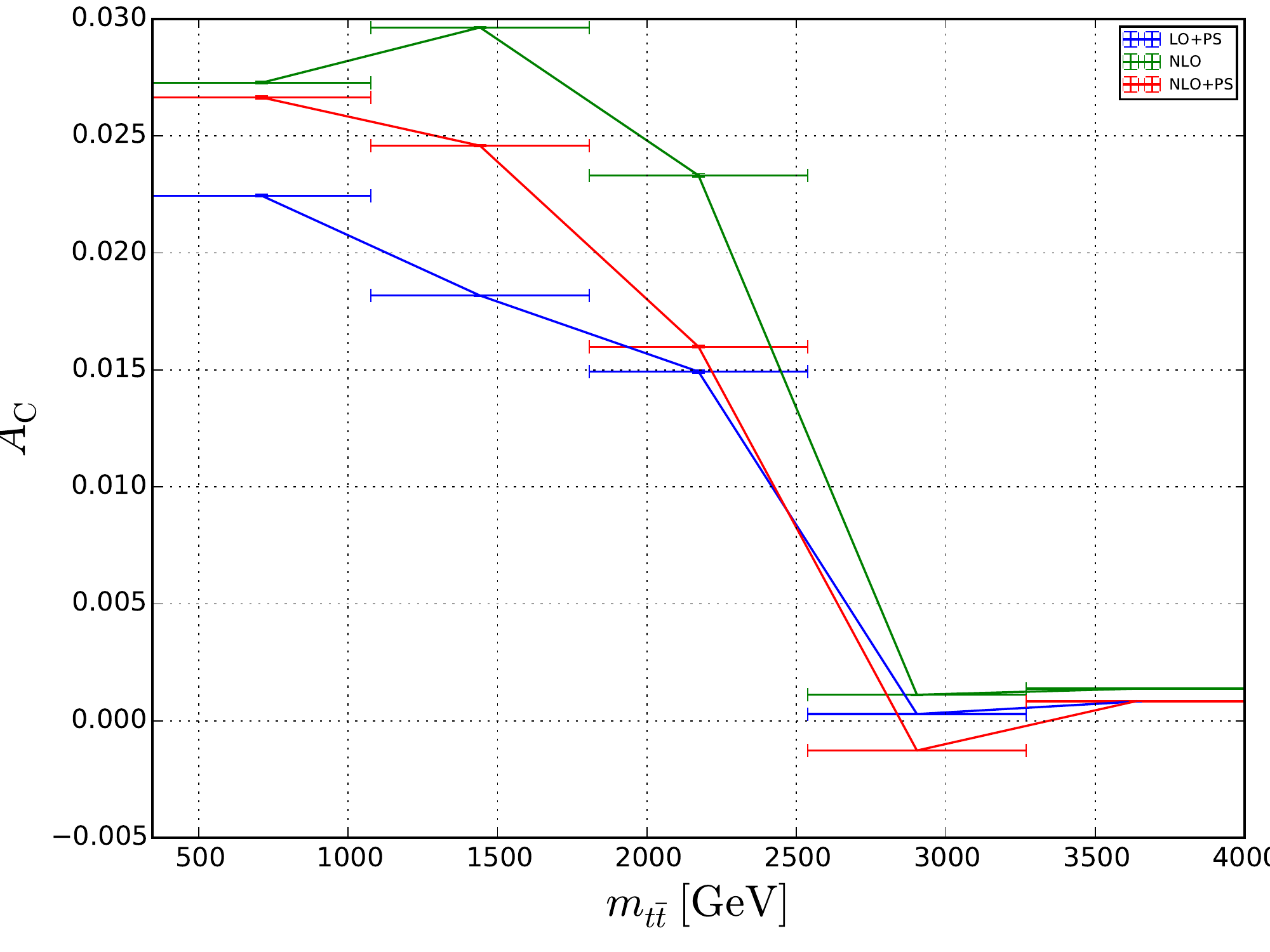}
 \caption{Invariant-mass distributions of the charge asymmetry $A_{C}$ of
 top-quark pairs produced through $\gamma$, $Z$ and
 $Z'$ bosons and their interferences at the LHC with $\sqrt{S}=13$ TeV at
 LO+PS (dark blue), NLO (green) and NLO+PS (red) accuracy together with the corresponding
 $K$-factors in the SSM (top) and TC (bottom) (color online).}
 \label{fig:15}
\end{figure}

A particularly sensitive observable for the distinction of new physics models is the
forward-backward asymmetry
\bea
 A_{FB}&=&{N(\Delta y>0)-N(\Delta y<0)
 \over
 N(\Delta y>0)+N(\Delta y<0)}
\eea
defined at $p\bar{p}$ colliders, where $\Delta y=y_t-y_{\bar{t}}$ is the rapidity difference of top
and antitop quarks, and the somewhat more complex charge asymmetry
\bea
 A_C&=&{N(\Delta |y|>0)-N(\Delta |y|<0)
 \over
 N(\Delta |y|>0)+N(\Delta |y|<0)}
\eea
defined at $pp$ colliders, where $\Delta |y|=|y_t|-|y_{\bar{t}}|$ is the corresponding difference
in absolute rapidity \cite{AguilarSaavedra:2012rx}. In Fig.\ \ref{fig:15}, the sensitivity of
$A_C$ to distinguish between the SSM (top) and TC (bottom) is confirmed, as this observable
exhibits very different magnitudes at the resonance ($11\pm1$\% vs.\ $\pm0.1$\%) and far below it
($2.5\pm0.5$\% in both plots), where the SM contributions dominate. Since $A_C$ is
defined as a ratio of cross sections, NLO and PS corrections cancel to a large extent and
are barely visible above the statistical noise. Only for TC, where the rapidity distribution
in Fig.\ \ref{fig:14} (lowest panel) showed distinct features in the ratio of NLO+PS/LO+PS,
the transition from the low-mass to the resonance region happens more abruptly in fixed order
(NLO) than with PS.
If we assume an integrated luminosity of 100 fb$^{-1}$ and integrate over an invariant-mass
window of 100 GeV around the resonance peak at 3 TeV, one would expect $10^{-5}$ pb/GeV$\times100$
fb$^{-1}\times100$ GeV
= 100 events. A 10\% asymmetry in the SSM then implies a difference of 10 events with an
error of 3, so that $A_C=(10\pm 3)\%$. This would be sufficient to distinguish the SSM
from the SM and TC.

\section{Conclusions}
\label{sec:5}

In this paper we presented the calculation of the ${\mathcal O}(\alpha_S\alpha^2)$
corrections to the electroweak production of top-antitop pairs through SM photons,
$Z$ and $Z'$ bosons, as predicted in the Sequential SM or in tecnicolor models.
Our corrections are implemented in the NLO parton shower Monte Carlo program
POWHEG. $Z'$ reconances are actively searched for by the ATLAS and CMS experiments
at the LHC with its now increased
center-of-mass energy of 13 TeV. We have consistently included interferences
between SM and new physics contributions and have introduced a proper subtraction
formalism for QED singularities. With a great variety of numerical predictions, we have
demonstrated the mass dependence of the $K$-factor, the changing relative sizes
of scale and PDF uncertainties, the large impact of new
partonic channels opening up at NLO (in particular of those induced by photon
PDFs in the proton), and the non-negligibility of interference effects.
Distributions in invariant mass were shown to be particularly sensitive to the
latter. The all-order resummation of perturbative corrections implicit in
the parton shower has been shown to make the transverse-momentum and azimuthal
angle distributions of the top-antitop pair finite and physical. Heavy new
gauge-boson contributions were seen to lead to much more centrally produced
top pairs, and the charge asymmetry has been shown to be a promising observable
to distinguish between different new physics models. Our implementation of this
new process in POWHEG, called PBZp, is very flexible, as it allows for the
simulation of any $Z'$-boson model, and should thus prove to be a useful tool
for $Z'$-boson searches in the top-antitop channel at the LHC, in particular
for leptophobic models.

\subsection*{Acknowledgments}

We thank J.\ Gao for making possible detailed numerical comparisons of our NLO
$Z'$ calculations as well as R.M.\ Harris and J.\ Ferrando for help with
matching their LO calculations in the topcolor model.
T.J.\ thanks P.\ Nason and C.\ Oleari for useful discussions.
This work was partially supported by the BMBF Verbundprojekt 05H2015 through grant 05H15PMCCA,
the DFG Graduiertenkolleg 2149, the CNRS/IN2P3 Theory-LHC-France initiative, and the EU program
FP7/2007-2013 through grant 302997.


\appendix
\clearpage

\section{Master integrals}
\label{sec:a}

The three master integrals needed for the calculation of our NLO corrections
are
the massive tadpole, $T(m^2)$, the massless two-point function, $B(0,0,p^2)$
and the
massive two-point function, $B(a,a,p^2)$. Their analytic expressions in
Laurent series
of $(D-4)$, up to $\mathcal{O}((D-4))$, are given in the Euclidean region by
the following formulas:
\bea
T(m^2) & = & \mu_0^{(4-D)} \int d^{D}k \frac{1}{(k^2+m^2)} \, , \nonumber \\
& = & \pi^{\frac{D}{2}} \Gamma \left( 3-\frac{D}{2} \right) \, \left(
\frac{m^2}{\mu_0^2}
\right)^{\frac{D-4}{2}} \! \! \! \! m^2 \biggl\{ \frac{2}{(D-4)} - 1 -
\frac{1}{2} (D-4) \nonumber \\
& & + \mathcal{O}( (D-4)^2 )
\biggr\} \, , \\
B(0,0,p^2) & = & \mu_0^{(4-D)} \int d^{D}k \frac{1}{k^2(p-k)^2} \, ,
\nonumber \\
& = & \pi^{\frac{D}{2}} \Gamma \left( 3-\frac{D}{2} \right) \, \left(
\frac{m^2}{\mu_0^2}
\right)^{\frac{D-4}{2}} \biggl\{ - \frac{2}{(D-4)} + \bigl[ 2 + H(0;x) + 2
H(1;x) \bigr] \nonumber \\
& & + (D-4) \biggl[ - 2 + \frac{1}{4}\zeta(2) -
    H(0;x) - \frac{1}{2} H(0,0;x) - H(0,1;x) \nonumber \\
& & \hspace*{17mm} -2 H(1;x) - H(1,0;x) - 2 H(1,1;x) \biggr] + \mathcal{O}(
(D-4)^2 ) \biggr\}
\, , \\
B(m^2,m^2,p^2) &=& \mu_0^{(4-D)} \int d^{D}k \frac{1}{(k^2+m^2)[(p-k)^2+m^2]}
\, , \nonumber \\
& = & \pi^{\frac{D}{2}} \Gamma \left( 3-\frac{D}{2} \right) \, \left(
\frac{m^2}{\mu_0^2}
\right)^{\frac{D-4}{2}} \biggl\{ - \frac{2}{(D-4)} + \biggl[ 2 -  H(0;x) +
\frac{2}{(1-x)} H(0;x)
\biggr] \nonumber \\
& & +(D-4) \biggl[ - 2 + \frac{1}{(1-x)}\zeta(2) -
\frac{1}{2}\zeta(2) -  H(-1,0;x)  + \frac{2}{(1-x)} H(-1,0;x)  \nonumber \\
& & \hspace*{17mm}  + H(0;x) - \frac{2}{(1-x)} H(0;x) +
\frac{1}{2} H(0,0;x) - \frac{1}{(1-x)} H(0,0;x) \biggr] \nonumber \\
& & + \mathcal{O}( (D-4)^2 ) \biggr\} \, ,
\eea
where $p^2 = m^2(1-x)^2/x$, $\zeta$ is the Riemann $\zeta$ function and where
the functions $H$
denote the harmonic polylogarithms (HPLs) of variable $x$
\cite{Remiddi:1999ew, Goncharov:1998kja}.

\section{Integrated dipole counter terms}
\label{sec:b}

The integrated dipole counter terms are obtained from the general expression  \cite{Catani:2002hc}
\begin{eqnarray}
  \bold{I}(\varepsilon,\mu_r^{2},{p_i,m_i}) &=& -\frac{\alpha_S}{2\pi}\frac{(4\pi)^{\varepsilon}}{\Gamma(1-\varepsilon)}\frac{1}{\bold{T}_{i}^2}\bold{T}_{i}\cdot \bold{T}_{j}\times \left[ \bold{T}_{i}^2 \left(  \frac{\mu_r^{2}}{s_{ij}} \right)^{\varepsilon} V_j(m_i,m_j,\varepsilon) +\Gamma_j(m_j,\varepsilon)\right]\qquad
  \label{eq:b.1}\\
  &+& i\leftrightarrow j+ \mathrm{finite\ terms}\,,\nonumber
\end{eqnarray}
where $\bold{T}^l$ denotes the color matrix associated with parton $l$ ($\bold{T}^{l}_{cb}=if_{clb}$
for gluons, $\bold{T}^{l}_{ab}=t^{l}_{ab}$ and $\bold{T}^{l}_{ab}=-t^{l}_{ba}$ for quarks and
anti-quarks), $s_{ij}=2p_i\cdot p_j$, and
\begin{align}
 V_j(0,0,\varepsilon) &=\frac{1}{\varepsilon^2}\,,\  &V_j(m_t,m_t,\varepsilon) &
 =\frac{1}{\varepsilon}\frac{1}{v_{ji}}\ln{\rho}\,, \label{eq:b.2} \\
 \Gamma_j(0,\varepsilon) &= \frac{\gamma_q}{\varepsilon}\,,\ &\Gamma_j(m_j,\varepsilon) &
 = \frac{C_F}{\varepsilon}\label{eq:b.3}
\end{align}
with $v_{ji}=\sqrt{1- \frac{p_j^2p_i^2}{(p_i\cdot p_j)^2}}$, $\rho=\sqrt{\frac{1-v_{ji}}{1+v_{ji}}}$
and $\gamma_q=3/2C_F$. Using Eqs.~(\ref{eq:b.1}), (\ref{eq:b.2}) and (\ref{eq:b.3}), we find
\begin{eqnarray}
  \bold{I}_{\mathrm{init.}} &=& \frac{2\alpha_S}{2\pi}\frac{(4\pi)^{\varepsilon}}{\Gamma(1-\varepsilon)}\left(\left( \frac{\mu_r^2}{s} \right)^{\varepsilon}\frac{C_F}{\varepsilon^{2}} +\frac{\gamma_q}{\varepsilon} \right)+\mathrm{finite\ terms}\\
  \bold{I}_{\mathrm{final}} &=& \frac{2\alpha_S}{2\pi}\frac{(4\pi)^{\varepsilon}}{\Gamma(1-\varepsilon)}\left( \frac{C_F}{\varepsilon}\left( \frac{\mu_r^2}{s-2m_t^2} \right)^{\varepsilon}\frac{1+x^2}{1-x^2}\ln{x} +\frac{C_F}{\varepsilon} \right)+\mathrm{finite\ terms}\,,
\end{eqnarray}
where again $s=m_t^2(1+x)^2/x$ and where the double poles are seen to originate only
from initial-state massless quarks.
The IR poles are given by the Born cross section multiplied by a factor
${\bf I}_{\rm init}+{\bf I}_{\rm final}$. 

\bibliographystyle{JHEP}
\bibliography{paper}

\end{document}